\pdfoutput=1
\PassOptionsToPackage{pdftex,
pdfencoding=auto,
pdfnewwindow=true,
pdfusetitle=true,
bookmarks=true,
bookmarksnumbered=true,
bookmarksopen=true,
pdfpagemode=UseThumbs,
bookmarksopenlevel=1,
pdfpagelabels=false,
breaklinks=true,
}{hyperref}
\documentclass[aps,pra,reprint,superscriptaddress,tightenlines,
longbibliography,nofootinbib,twocolumn,letterpaper,notitlepage
,floatfix,noeprint]{revtex4-2}
\PassOptionsToPackage{pdftex,
pdfencoding=auto,
pdfnewwindow=true,
pdfusetitle=true,
bookmarks=true,
bookmarksnumbered=true,
bookmarksopen=true,
pdfpagemode=UseThumbs,
bookmarksopenlevel=1,
pdfpagelabels=false,
breaklinks=true,
}{hyperref}
\usepackage[utf8]{inputenx}
\usepackage{textcomp}
\usepackage{lmodern}
\usepackage{microtype}
\microtypesetup{
expansion={true,nocompatibility},
protrusion={true,nocompatibility},
activate={true,nocompatibility},
tracking=true,
kerning=true,
spacing=false
}
\usepackage[subtle,title,mathdisplays=normal,indent=normal,tracking=normal,paragraphs=normal]{savetrees}
\usepackage[style=american,autopunct=true]{csquotes}

\usepackage{amsthm,amsmath,amssymb,comment}
\usepackage{bbold}
\usepackage{hyphenat}
\usepackage{units}
\usepackage{multirow}
\usepackage[normalem]{ulem}
\usepackage{dsfont}
\usepackage{xspace,enumerate}
\usepackage[pdftex]{graphicx}
\usepackage{marvosym}
\usepackage{ulem}

\usepackage{tikzfig}
\pgfplotsset{compat=newest}

\usepackage{scalerel}
\usepackage{enumitem}
\usepackage{bm}
\usepackage[]{xcolor} 
\usepackage{placeins} 
\definecolor{purple}{RGB}{128,0,128}
\definecolor{ultramarine}{RGB}{63, 0, 255}
\definecolor{medblue}{RGB}{0, 0, 100}
\definecolor{panblue}{RGB}{0,24,150}
\definecolor{googleblue}{RGB}{34, 0, 204}
\definecolor{carmine}{RGB}{150, 0, 24}
\definecolor{gray}{RGB}{150, 150, 150}
\definecolor{darkred}{RGB}{200, 0, 0}
\definecolor{darkgreen}{RGB}{0, 80, 0}
\definecolor{medgreen}{RGB}{1, 80, 32}
\definecolor{darkblue}{RGB}{0, 0, 200}
\definecolor{nred}{rgb}{0.9,0.1,0.1}
\definecolor{nblack}{rgb}{0,0,0}
\definecolor{nblue}{rgb}{0.2,0.2,0.8}
\definecolor{ngreen}{rgb}{0.2,0.6,0.2}

\usepackage[unicode=true,pdfusetitle, bookmarks=false,bookmarksnumbered=false,
bookmarksopen=false, breaklinks=false,pdfborder={0 0 0},backref=false,
colorlinks=true,
linkcolor=carmine,
citecolor=googleblue,
urlcolor=panblue,
anchorcolor=medgreen
]{hyperref}

\newcommand{\blk}{\color{nblack}}

\newcommand{\st}{\ensuremath{\Omega}}

\newcommand{\GPTt}{\ensuremath{\Omega}}
\newcommand{\gpEff}{\ensuremath{\mathcal{E}}}
\newcommand{\gp}[1]{\ensuremath{\mathbf{#1}}}

\newcommand{\ketbra}[2]{\ensuremath{{\textstyle{\ket{#1}}\!\!\!\!{\bra{#2}}}}}

\usepackage{braket}

\newcommand\restr[2]{{
  \left.\kern-\nulldelimiterspace 
  #1 
  \vphantom{\big|} 
  \right|_{#2} 
  }}
\newcommand{\bit}{\begin{itemize}}
\newcommand{\eit}{\end{itemize}\par\noindent}
\newcommand{\ben}{\begin{enumerate}}
\newcommand{\een}{\end{enumerate}\par\noindent}
\newcommand{\beq}{\begin{equation}}
\newcommand{\eeq}{\end{equation}\par\noindent}
\newcommand{\beqa}{\begin{eqnarray*}}
\newcommand{\eeqa}{\end{eqnarray*}\par\noindent}
\newcommand{\beqn}{\begin{eqnarray}}
\newcommand{\eeqn}{\end{eqnarray}\par\noindent}

\newtheorem{theoremNew}{Theorem}

\newtheorem{definitionNew}{Definition}

\newtheorem{proposition}{Proposition}

\usetikzlibrary{shapes.multipart}

\tikzstyle{arrowhead}=[regular polygon,regular polygon sides=3,draw,scale=0.2,inner sep=-0.15pt,minimum width=6mm,fill=black,regular polygon rotate=180]
\tikzstyle{trace}=[circuit ee IEC,thick,ground,rotate=0,scale=2]
\tikzstyle{wavy}=[decorate,decoration={snake, segment length=1mm, amplitude=0.3mm}]
\tikzstyle{mopoint}=[shape=semicircle, fill=white,draw=black,shape border rotate=180,scale =0.75]
\tikzstyle{mocopoint}=[shape=semicircle, fill=white,draw=black,minimum width = 0.9cm, scale =0.75, xscale=0.7]
\tikzstyle{cpoint}=[shape=semicircle, fill=white,draw=black,minimum width = 0.9cm, scale =0.75, xscale=1, yscale=0.7, shape border rotate = 90,font=\fontsize{14}{16}\selectfont]
\tikzstyle{cocpoint}=[shape=semicircle, fill=white,draw=black,minimum width = 0.9cm, scale =0.75, xscale=1, yscale=0.7, shape border rotate = 270,font=\fontsize{14}{16}\selectfont]
\tikzstyle{slit}=[line width=2]
\tikzstyle{block}=[line width=4,red,line cap=round]
\tikzstyle{screen}=[line width=4,black,line cap=round]
\tikzstyle{di}=[diamond,draw,inner sep=0.5pt,font=\small, minimum size = .5cm]
\tikzstyle{sbox}=[rectangle,draw]
\tikzstyle{mirror}=[line width=2,black]
\tikzstyle{traceState}=[circuit ee IEC,thick,ground,rotate=180,scale=2]
\tikzstyle{detEff}=[circuit ee IEC,thick,ground,rotate=180,scale=1.4]
\tikzstyle{maxMix}=[circuit ee IEC,thick,ground,scale=1.4]
\tikzstyle{particlePath}=[line width=2,gray!40, line cap =round]

\tikzstyle{bwSpider}=[
       rectangle split,
       rectangle split parts=2,
       rectangle split part fill={black,white},
 minimum size=3.6 mm, inner sep=-2mm, draw=black,scale=0.5,rounded corners=0.8 mm
       ]
 \tikzstyle{wbSpider}=[
       rectangle split,
       rectangle split parts=2,
       rectangle split part fill={white,black},
 minimum size=3.6 mm, inner sep=-2mm, draw=black,scale=0.5,rounded corners=0.8 mm
       ]
\tikzstyle{cWire}=[densely dotted, thick]
\tikzstyle{env}=[copoint,regular polygon rotate=0,minimum width=0.2cm, fill=black]

\tikzstyle{probs}=[shape=semicircle,fill=white,draw=black,shape border rotate=180,minimum width=1.2cm]

\tikzstyle{every picture}=[baseline=-0.25em,scale=0.5]
\tikzstyle{dotpic}=[] 
\tikzstyle{diredges}=[every to/.style={diredge}]
\tikzstyle{math matrix}=[matrix of math nodes,left delimiter=(,right delimiter=),inner sep=2pt,column sep=1em,row sep=0.5em,nodes={inner sep=0pt},text height=1.5ex, text depth=0.25ex]

\tikzstyle{inline text}=[text height=1.5ex, text depth=0.25ex,yshift=0.5mm]
\tikzstyle{label}=[font=\footnotesize,text height=1.5ex, text depth=0.25ex,yshift=0.5mm]
\tikzstyle{left label}=[label,anchor=east,xshift=1.mm]
\tikzstyle{right label}=[label,anchor=west,xshift=-1.mm]

\tikzstyle{braceedge}=[decorate,decoration={brace,amplitude=2mm,raise=-1mm}]
\tikzstyle{small braceedge}=[decorate,decoration={brace,amplitude=1mm,raise=-1mm}]

\tikzstyle{doubled}=[line width=1.6pt] 
\tikzstyle{boldedge}=[doubled,shorten <=-0.17mm,shorten >=-0.17mm]
\tikzstyle{boldedgegray}=[doubled,gray,shorten <=-0.17mm,shorten >=-0.17mm]
\tikzstyle{singleedgegray}=[gray]

\tikzstyle{semidoubled}=[line width=1.4pt] 
\tikzstyle{semiboldedgegray}=[semidoubled,gray,shorten <=-0.17mm,shorten >=-0.17mm]

\tikzstyle{boxedge}=[semiboldedgegray]

\tikzstyle{boldedgedashed}=[very thick,dashed,shorten <=-0.17mm,shorten >=-0.17mm]
\tikzstyle{vboldedgedashed}=[doubled,dashed,shorten <=-0.17mm,shorten >=-0.17mm]
\tikzstyle{left hook arrow}=[left hook-latex]
\tikzstyle{right hook arrow}=[right hook-latex]
\tikzstyle{sembracket}=[line width=0.5pt,shorten <=-0.07mm,shorten >=-0.07mm]

\tikzstyle{causal edge}=[->,thick,gray]
\tikzstyle{causal nondir}=[thick,gray]
\tikzstyle{timeline}=[thick,gray, dashed]

\tikzstyle{cedge}=[<->,thick,gray!70!white]

\tikzstyle{empty diagram}=[draw=gray!40!white,dashed,shape=rectangle,minimum width=1cm,minimum height=1cm]
\tikzstyle{empty diagram small}=[draw=gray!50!white,dashed,shape=rectangle,minimum width=0.6cm,minimum height=0.5cm]

\tikzstyle{dot}=[inner sep=0mm,minimum width=2mm,minimum height=2mm,draw,shape=circle]

\tikzstyle{leak}=[white dot, shape=regular polygon, minimum size=3.3 mm, regular polygon sides=3, outer sep=-0.2mm, regular polygon rotate=270]
\tikzstyle{proj}=[white dot, shape=regular polygon, minimum size=3.3 mm, regular polygon sides=4, outer sep=-0.2mm]
\tikzstyle{Vleak}=[white dot, shape=regular polygon, minimum size=3.3 mm, regular polygon sides=3, outer sep=-0.2mm, regular polygon rotate=90]
\tikzstyle{dleak}=[white dot, line width=1.6pt, shape=regular polygon, minimum size=3.3 mm, regular polygon sides=3, outer sep=-0.2mm, regular polygon rotate=270]

\tikzstyle{Wsquare}=[white dot, shape=regular polygon, rounded corners=0.8 mm, minimum size=3.3 mm, regular polygon sides=3, outer sep=-0.2mm]
\tikzstyle{Wsquareadj}=[white dot, shape=regular polygon, rounded corners=0.8 mm, minimum size=3.3 mm, regular polygon sides=3, outer sep=-0.2mm, regular polygon rotate=180]
\tikzstyle{ddot}=[inner sep=0mm, doubled, minimum width=2.5mm,minimum height=2.5mm,draw,shape=circle]
\tikzstyle{blue dot}=[dot,fill=blue,draw=blue]
\tikzstyle{black dot}=[dot,fill=black]
\tikzstyle{white dot}=[dot,fill=white,,text depth=-0.2mm]
\tikzstyle{white Wsquare}=[Wsquare,fill=gray,,text depth=-0.2mm]
\tikzstyle{white Wsquareadj}=[Wsquareadj,fill=white,,text depth=-0.2mm]
\tikzstyle{green dot}=[white dot] 
\tikzstyle{gray dot}=[dot,fill=gray!40!white,,text depth=-0.2mm]
\tikzstyle{red dot}=[gray dot] 

\tikzstyle{black ddot}=[ddot,fill=black]
\tikzstyle{white ddot}=[ddot,fill=white]
\tikzstyle{gray ddot}=[ddot,fill=gray!40!white]

\tikzstyle{gray edge}=[gray!60!white]

\tikzstyle{small dot}=[inner sep=0.5mm,minimum width=0pt,minimum height=0pt,draw,shape=circle]

\tikzstyle{small black dot}=[small dot,fill=black]
\tikzstyle{small white dot}=[small dot,fill=white]
\tikzstyle{small gray dot}=[small dot,fill=gray!40!white]

\tikzstyle{causal dot}=[inner sep=0.4mm,minimum width=0pt,minimum height=0pt,draw=white,shape=circle,fill=gray!40!white]

\tikzstyle{phase dimensions}=[minimum size=5mm,font=\footnotesize,rectangle,rounded corners=2.5mm,inner sep=0.2mm,outer sep=-2mm]
\tikzstyle{dphase dimensions}=[minimum size=5mm,font=\footnotesize,rectangle,rounded corners=2.5mm,inner sep=0.2mm,outer sep=-2mm]

\tikzstyle{white phase dot}=[dot,fill=white,phase dimensions]
\tikzstyle{white phase ddot}=[ddot,fill=white,dphase dimensions]

\tikzstyle{white rect ddot}=[draw=black,fill=white,doubled,minimum size=5mm,font=\footnotesize,rectangle,rounded corners=2.5mm,inner sep=0.2mm]
\tikzstyle{gray rect ddot}=[draw=black,fill=gray!40!white,doubled,minimum size=6mm,font=\footnotesize,rectangle,rounded corners=3mm]

\tikzstyle{gray phase dot}=[dot,fill=gray!40!white,phase dimensions]
\tikzstyle{gray phase ddot}=[ddot,fill=gray!40!white,dphase dimensions]
\tikzstyle{grey phase dot}=[gray phase dot]
\tikzstyle{grey phase ddot}=[gray phase ddot]

\tikzstyle{small phase dimensions}=[minimum size=4mm,font=\tiny,rectangle,rounded corners=2mm,inner sep=0.2mm,outer sep=-2mm]
\tikzstyle{small dphase dimensions}=[minimum size=4mm,font=\tiny,rectangle,rounded corners=2mm,inner sep=0.2mm,outer sep=-2mm]

\tikzstyle{small gray phase dot}=[dot,fill=gray!40!white,small phase dimensions]
\tikzstyle{small gray phase ddot}=[ddot,fill=gray!40!white,small dphase dimensions]

\tikzstyle{small map}=[draw,shape=rectangle,minimum height=4mm,minimum width=4mm,fill=white]

\tikzstyle{cnot}=[fill=white,shape=circle,inner sep=-1.4pt]

\tikzstyle{asym hadamard}=[fill=white,draw,shape=NEbox,inner sep=0.6mm,font=\footnotesize,minimum height=4mm]
\tikzstyle{asym hadamard conj}=[fill=white,draw,shape=NWbox,inner sep=0.6mm,font=\footnotesize,minimum height=4mm]
\tikzstyle{asym hadamard dag}=[fill=white,draw,shape=SEbox,inner sep=0.6mm,font=\footnotesize,minimum height=4mm]

\tikzstyle{hadamard}=[fill=white,draw,inner sep=0.6mm,font=\footnotesize,minimum height=4mm,minimum width=4mm]
\tikzstyle{small hadamard}=[fill=white,draw,inner sep=0.6mm,minimum height=1.5mm,minimum width=1.5mm]
\tikzstyle{small hadamard rotate}=[small hadamard,rotate=45]
\tikzstyle{dhadamard}=[hadamard,doubled]
\tikzstyle{small dhadamard}=[small hadamard,doubled]
\tikzstyle{small dhadamard rotate}=[small hadamard rotate,doubled]
\tikzstyle{antipode}=[white dot,inner sep=0.3mm,font=\footnotesize]

\tikzstyle{scalar}=[diamond,draw,inner sep=0.5pt,font=\small]
\tikzstyle{dscalar}=[diamond,doubled, draw,inner sep=0.5pt,font=\small]

\tikzstyle{small box}=[rectangle,inline text,fill=white,draw,minimum height=5mm,yshift=-0.5mm,minimum width=5mm,font=\small]
\tikzstyle{small gray box}=[small box,fill=gray!30]
\tikzstyle{medium box}=[rectangle,inline text,fill=white,draw,minimum height=5mm,yshift=-0.5mm,minimum width=10mm,font=\small]
\tikzstyle{square box}=[small box] 
\tikzstyle{medium gray box}=[small box,fill=gray!30]
\tikzstyle{semilarge box}=[rectangle,inline text,fill=white,draw,minimum height=5mm,yshift=-0.5mm,minimum width=12.5mm,font=\small]
\tikzstyle{large box}=[rectangle,inline text,fill=white,draw,minimum height=5mm,yshift=-0.5mm,minimum width=15mm,font=\small]
\tikzstyle{large gray box}=[small box,fill=gray!30]

\tikzstyle{Bayes box}=[rectangle,fill=black,draw, minimum height=3mm, minimum width=3mm]

\tikzstyle{gray square point}=[small box,fill=gray!50]

\tikzstyle{dphase box white}=[dhadamard]
\tikzstyle{dphase box gray}=[dhadamard,fill=gray!50!white]
\tikzstyle{phase box white}=[hadamard]
\tikzstyle{phase box gray}=[hadamard,fill=gray!50!white]

\tikzstyle{point}=[regular polygon,regular polygon sides=3,draw,scale=0.75,inner sep=-0.5pt,minimum width=9mm,fill=white,regular polygon rotate=180]
\tikzstyle{point nosep}=[regular polygon,regular polygon sides=3,draw,scale=0.75,inner sep=-2pt,minimum width=9mm,fill=white,regular polygon rotate=180]
\tikzstyle{copoint}=[regular polygon,regular polygon sides=3,draw,scale=0.75,inner sep=-0.5pt,minimum width=9mm,fill=white]
\tikzstyle{dpoint}=[point,doubled]
\tikzstyle{dcopoint}=[copoint,doubled]

\tikzstyle{pointgrow}=[shape=cornerpoint,kpoint common,scale=0.75,inner sep=3pt]
\tikzstyle{pointgrow dag}=[shape=cornercopoint,kpoint common,scale=0.75,inner sep=3pt]

\tikzstyle{wide copoint}=[fill=white,draw,shape=isosceles triangle,shape border rotate=90,isosceles triangle stretches=true,inner sep=0pt,minimum width=1.5cm,minimum height=6.12mm]
\tikzstyle{wide point}=[fill=white,draw,shape=isosceles triangle,shape border rotate=-90,isosceles triangle stretches=true,inner sep=0pt,minimum width=1.5cm,minimum height=6.12mm,yshift=-0.0mm]
\tikzstyle{wide point plus}=[fill=white,draw,shape=isosceles triangle,shape border rotate=-90,isosceles triangle stretches=true,inner sep=0pt,minimum width=1.74cm,minimum height=7mm,yshift=-0.0mm]

\tikzstyle{wide dpoint}=[fill=white,doubled,draw,shape=isosceles triangle,shape border rotate=-90,isosceles triangle stretches=true,inner sep=0pt,minimum width=1.5cm,minimum height=6.12mm,yshift=-0.0mm]

\tikzstyle{tinypoint}=[regular polygon,regular polygon sides=3,draw,scale=0.55,inner sep=-0.15pt,minimum width=6mm,fill=white,regular polygon rotate=180]

\tikzstyle{white point}=[point]
\tikzstyle{white dpoint}=[dpoint]
\tikzstyle{green point}=[white point] 
\tikzstyle{white copoint}=[copoint]
\tikzstyle{gray point}=[point,fill=gray!40!white]
\tikzstyle{gray dpoint}=[gray point,doubled]
\tikzstyle{red point}=[gray point] 
\tikzstyle{gray copoint}=[copoint,fill=gray!40!white]
\tikzstyle{gray dcopoint}=[gray copoint,doubled]

\tikzstyle{white point guide}=[regular polygon,regular polygon sides=3,font=\scriptsize,draw,scale=0.65,inner sep=-0.5pt,minimum width=9mm,fill=white,regular polygon rotate=180]

\tikzstyle{black point}=[point,fill=black,font=\color{white}]
\tikzstyle{black copoint}=[copoint,fill=black,font=\color{white}]

\tikzstyle{tiny gray point}=[tinypoint,fill=gray!40!white]

\tikzstyle{diredge}=[->]
\tikzstyle{ddiredge}=[<->]
\tikzstyle{rdiredge}=[<-]
\tikzstyle{thickdiredge}=[->, very thick]
\tikzstyle{pointer edge}=[->,very thick,gray]
\tikzstyle{pointer edge part}=[very thick,gray]
\tikzstyle{dashed edge}=[dashed]
\tikzstyle{thick dashed edge}=[very thick,dashed]
\tikzstyle{thick gray dashed edge}=[thick dashed edge,gray!40]
\tikzstyle{thick blue dashed edge}=[thick dashed edge,blue!40]
\tikzstyle{thick map edge}=[very thick,|->]

\makeatletter
\newcommand{\boxshape}[3]{%
\pgfdeclareshape{#1}{
\inheritsavedanchors[from=rectangle] 
\inheritanchorborder[from=rectangle]
\inheritanchor[from=rectangle]{center}
\inheritanchor[from=rectangle]{north}
\inheritanchor[from=rectangle]{south}
\inheritanchor[from=rectangle]{west}
\inheritanchor[from=rectangle]{east}
\backgroundpath{
\southwest \pgf@xa=\pgf@x \pgf@ya=\pgf@y
\northeast \pgf@xb=\pgf@x \pgf@yb=\pgf@y

\@tempdima=#2
\@tempdimb=#3

\pgfpathmoveto{\pgfpoint{\pgf@xa - 5pt + \@tempdima}{\pgf@ya}}
\pgfpathlineto{\pgfpoint{\pgf@xa - 5pt - \@tempdima}{\pgf@yb}}
\pgfpathlineto{\pgfpoint{\pgf@xb + 5pt + \@tempdimb}{\pgf@yb}}
\pgfpathlineto{\pgfpoint{\pgf@xb + 5pt - \@tempdimb}{\pgf@ya}}
\pgfpathlineto{\pgfpoint{\pgf@xa - 5pt + \@tempdima}{\pgf@ya}}
\pgfpathclose
}
}}

\boxshape{NEbox}{0pt}{5pt}
\boxshape{SEbox}{0pt}{-5pt}
\boxshape{NWbox}{5pt}{0pt}
\boxshape{SWbox}{-5pt}{0pt}
\boxshape{EBox}{-3pt}{3pt}
\boxshape{WBox}{3pt}{-3pt}
\makeatother

\tikzstyle{cloud}=[shape=cloud,draw,minimum width=1.5cm,minimum height=1.5cm]

\tikzstyle{map}=[draw,shape=NEbox,inner sep=2pt,minimum height=6mm,fill=white]
\tikzstyle{dashedmap}=[draw,dashed,shape=NEbox,inner sep=2pt,minimum height=6mm,fill=white]
\tikzstyle{mapdag}=[draw,shape=SEbox,inner sep=2pt,minimum height=6mm,fill=white]
\tikzstyle{mapadj}=[draw,shape=SEbox,inner sep=2pt,minimum height=6mm,fill=white]
\tikzstyle{maptrans}=[draw,shape=SWbox,inner sep=2pt,minimum height=6mm,fill=white]
\tikzstyle{mapconj}=[draw,shape=NWbox,inner sep=2pt,minimum height=6mm,fill=white]

\tikzstyle{medium map}=[draw,shape=NEbox,inner sep=2pt,minimum height=6mm,fill=white,minimum width=7mm]
\tikzstyle{medium map dag}=[draw,shape=SEbox,inner sep=2pt,minimum height=6mm,fill=white,minimum width=7mm]
\tikzstyle{medium map adj}=[draw,shape=SEbox,inner sep=2pt,minimum height=6mm,fill=white,minimum width=7mm]
\tikzstyle{medium map trans}=[draw,shape=SWbox,inner sep=2pt,minimum height=6mm,fill=white,minimum width=7mm]
\tikzstyle{medium map conj}=[draw,shape=NWbox,inner sep=2pt,minimum height=6mm,fill=white,minimum width=7mm]
\tikzstyle{semilarge map}=[draw,shape=NEbox,inner sep=2pt,minimum height=6mm,fill=white,minimum width=9.5mm]
\tikzstyle{semilarge map trans}=[draw,shape=SWbox,inner sep=2pt,minimum height=6mm,fill=white,minimum width=9.5mm]
\tikzstyle{semilarge map adj}=[draw,shape=SEbox,inner sep=2pt,minimum height=6mm,fill=white,minimum width=9.5mm]
\tikzstyle{semilarge map dag}=[draw,shape=SEbox,inner sep=2pt,minimum height=6mm,fill=white,minimum width=9.5mm]
\tikzstyle{semilarge map conj}=[draw,shape=NWbox,inner sep=2pt,minimum height=6mm,fill=white,minimum width=9.5mm]
\tikzstyle{large map}=[draw,shape=NEbox,inner sep=2pt,minimum height=6mm,fill=white,minimum width=12mm]
\tikzstyle{large map conj}=[draw,shape=NWbox,inner sep=2pt,minimum height=6mm,fill=white,minimum width=12mm]
\tikzstyle{very large map}=[draw,shape=NEbox,inner sep=2pt,minimum height=6mm,fill=white,minimum width=17mm]

\tikzstyle{medium dmap}=[draw,doubled,shape=NEbox,inner sep=2pt,minimum height=6mm,fill=white,minimum width=7mm]
\tikzstyle{medium dmap dag}=[draw,doubled,shape=SEbox,inner sep=2pt,minimum height=6mm,fill=white,minimum width=7mm]
\tikzstyle{medium dmap adj}=[draw,doubled,shape=SEbox,inner sep=2pt,minimum height=6mm,fill=white,minimum width=7mm]
\tikzstyle{medium dmap trans}=[draw,doubled,shape=SWbox,inner sep=2pt,minimum height=6mm,fill=white,minimum width=7mm]
\tikzstyle{medium dmap conj}=[draw,doubled,shape=NWbox,inner sep=2pt,minimum height=6mm,fill=white,minimum width=7mm]
\tikzstyle{semilarge dmap}=[draw,doubled,shape=NEbox,inner sep=2pt,minimum height=6mm,fill=white,minimum width=9.5mm]
\tikzstyle{semilarge dmap trans}=[draw,doubled,shape=SWbox,inner sep=2pt,minimum height=6mm,fill=white,minimum width=9.5mm]
\tikzstyle{semilarge dmap adj}=[draw,doubled,shape=SEbox,inner sep=2pt,minimum height=6mm,fill=white,minimum width=9.5mm]
\tikzstyle{semilarge dmap dag}=[draw,doubled,shape=SEbox,inner sep=2pt,minimum height=6mm,fill=white,minimum width=9.5mm]
\tikzstyle{semilarge dmap conj}=[draw,doubled,shape=NWbox,inner sep=2pt,minimum height=6mm,fill=white,minimum width=9.5mm]
\tikzstyle{large dmap}=[draw,doubled,shape=NEbox,inner sep=2pt,minimum height=6mm,fill=white,minimum width=12mm]
\tikzstyle{large dmap conj}=[draw,doubled,shape=NWbox,inner sep=2pt,minimum height=6mm,fill=white,minimum width=12mm]
\tikzstyle{large dmap trans}=[draw,doubled,shape=SWbox,inner sep=2pt,minimum height=6mm,fill=white,minimum width=12mm]
\tikzstyle{large dmap adj}=[draw,doubled,shape=SEbox,inner sep=2pt,minimum height=6mm,fill=white,minimum width=12mm]
\tikzstyle{large dmap dag}=[draw,doubled,shape=SEbox,inner sep=2pt,minimum height=6mm,fill=white,minimum width=12mm]
\tikzstyle{very large dmap}=[draw,doubled,shape=NEbox,inner sep=2pt,minimum height=6mm,fill=white,minimum width=19.5mm]

\tikzstyle{muxbox}=[draw,shape=rectangle,minimum height=3mm,minimum width=3mm,fill=white]
\tikzstyle{dmuxbox}=[muxbox,doubled]

\tikzstyle{box}=[draw,shape=rectangle,inner sep=2pt,minimum height=6mm,minimum width=6mm,fill=white]
\tikzstyle{dbox}=[draw,doubled,shape=rectangle,inner sep=2pt,minimum height=6mm,minimum width=6mm,fill=white]
\tikzstyle{dmap}=[draw,doubled,shape=NEbox,inner sep=2pt,minimum height=6mm,fill=white]
\tikzstyle{dmapdag}=[draw,doubled,shape=SEbox,inner sep=2pt,minimum height=6mm,fill=white]
\tikzstyle{dmapadj}=[draw,doubled,shape=SEbox,inner sep=2pt,minimum height=6mm,fill=white]
\tikzstyle{dmaptrans}=[draw,doubled,shape=SWbox,inner sep=2pt,minimum height=6mm,fill=white]
\tikzstyle{dmapconj}=[draw,doubled,shape=NWbox,inner sep=2pt,minimum height=6mm,fill=white]

\tikzstyle{ddmap}=[draw,doubled,dashed,shape=NEbox,inner sep=2pt,minimum height=6mm,fill=white]
\tikzstyle{ddmapdag}=[draw,doubled,dashed,shape=SEbox,inner sep=2pt,minimum height=6mm,fill=white]
\tikzstyle{ddmapadj}=[draw,doubled,dashed,shape=SEbox,inner sep=2pt,minimum height=6mm,fill=white]
\tikzstyle{ddmaptrans}=[draw,doubled,dashed,shape=SWbox,inner sep=2pt,minimum height=6mm,fill=white]
\tikzstyle{ddmapconj}=[draw,doubled,dashed,shape=NWbox,inner sep=2pt,minimum height=6mm,fill=white]

\boxshape{sNEbox}{0pt}{3pt}
\boxshape{sSEbox}{0pt}{-3pt}
\boxshape{sNWbox}{3pt}{0pt}
\boxshape{sSWbox}{-3pt}{0pt}
\tikzstyle{smap}=[draw,shape=sNEbox,fill=white]
\tikzstyle{smapdag}=[draw,shape=sSEbox,fill=white]
\tikzstyle{smapadj}=[draw,shape=sSEbox,fill=white]
\tikzstyle{smaptrans}=[draw,shape=sSWbox,fill=white]
\tikzstyle{smapconj}=[draw,shape=sNWbox,fill=white]

\tikzstyle{dsmap}=[draw,dashed,shape=sNEbox,fill=white]
\tikzstyle{dsmapdag}=[draw,dashed,shape=sSEbox,fill=white]
\tikzstyle{dsmaptrans}=[draw,dashed,shape=sSWbox,fill=white]
\tikzstyle{dsmapconj}=[draw,dashed,shape=sNWbox,fill=white]

\boxshape{mNEbox}{0pt}{10pt}
\boxshape{mSEbox}{0pt}{-10pt}
\boxshape{mNWbox}{10pt}{0pt}
\boxshape{mSWbox}{-10pt}{0pt}
\tikzstyle{mmap}=[draw,shape=mNEbox]
\tikzstyle{mmapdag}=[draw,shape=mSEbox]
\tikzstyle{mmaptrans}=[draw,shape=mSWbox]
\tikzstyle{mmapconj}=[draw,shape=mNWbox]

\tikzstyle{mmapgray}=[draw,fill=gray!40!white,shape=mNEbox]
\tikzstyle{smapgray}=[draw,fill=gray!40!white,shape=sNEbox]

\makeatletter

\pgfdeclareshape{cornerpoint}{
\inheritsavedanchors[from=rectangle] 
\inheritanchorborder[from=rectangle]
\inheritanchor[from=rectangle]{center}
\inheritanchor[from=rectangle]{north}
\inheritanchor[from=rectangle]{south}
\inheritanchor[from=rectangle]{west}
\inheritanchor[from=rectangle]{east}
\backgroundpath{
\southwest \pgf@xa=\pgf@x \pgf@ya=\pgf@y
\northeast \pgf@xb=\pgf@x \pgf@yb=\pgf@y

\pgfmathsetmacro{\pgf@shorten@left}{\pgfkeysvalueof{/tikz/shorten left}}
\pgfmathsetmacro{\pgf@shorten@right}{\pgfkeysvalueof{/tikz/shorten right}}

\pgfpathmoveto{\pgfpoint{0.5 * (\pgf@xa + \pgf@xb)}{\pgf@ya - 5pt}}
\pgfpathlineto{\pgfpoint{\pgf@xa - 8pt + \pgf@shorten@left}{\pgf@yb - 1.5 * \pgf@shorten@left}}
\pgfpathlineto{\pgfpoint{\pgf@xa - 8pt + \pgf@shorten@left}{\pgf@yb}}
\pgfpathlineto{\pgfpoint{\pgf@xb + 8pt - \pgf@shorten@right}{\pgf@yb}}
\pgfpathlineto{\pgfpoint{\pgf@xb + 8pt - \pgf@shorten@right}{\pgf@yb - 1.5 * \pgf@shorten@right}}
\pgfpathclose
}
}

\pgfdeclareshape{cornercopoint}{
\inheritsavedanchors[from=rectangle] 
\inheritanchorborder[from=rectangle]
\inheritanchor[from=rectangle]{center}
\inheritanchor[from=rectangle]{north}
\inheritanchor[from=rectangle]{south}
\inheritanchor[from=rectangle]{west}
\inheritanchor[from=rectangle]{east}
\backgroundpath{
\southwest \pgf@xa=\pgf@x \pgf@ya=\pgf@y
\northeast \pgf@xb=\pgf@x \pgf@yb=\pgf@y

\pgfmathsetmacro{\pgf@shorten@left}{\pgfkeysvalueof{/tikz/shorten left}}
\pgfmathsetmacro{\pgf@shorten@right}{\pgfkeysvalueof{/tikz/shorten right}}

\pgfpathmoveto{\pgfpoint{0.5 * (\pgf@xa + \pgf@xb)}{\pgf@yb + 5pt}}
\pgfpathlineto{\pgfpoint{\pgf@xa - 8pt + \pgf@shorten@left}{\pgf@ya + 1.5 * \pgf@shorten@left}}
\pgfpathlineto{\pgfpoint{\pgf@xa - 8pt + \pgf@shorten@left}{\pgf@ya}}
\pgfpathlineto{\pgfpoint{\pgf@xb + 8pt - \pgf@shorten@right}{\pgf@ya}}
\pgfpathlineto{\pgfpoint{\pgf@xb + 8pt - \pgf@shorten@right}{\pgf@ya + 1.5 * \pgf@shorten@right}}
\pgfpathclose
}
}

\makeatother

\pgfkeyssetvalue{/tikz/shorten left}{0pt}
\pgfkeyssetvalue{/tikz/shorten right}{0pt}

\tikzstyle{kpoint common}=[draw,fill=white,inner sep=1pt,minimum height=4mm]
\tikzstyle{kpoint sc}=[shape=cornerpoint,kpoint common]
\tikzstyle{kpoint adjoint sc}=[shape=cornercopoint,kpoint common]
\tikzstyle{kpoint}=[shape=cornerpoint,shorten left=5pt,kpoint common]
\tikzstyle{kpoint adjoint}=[shape=cornercopoint,shorten left=5pt,kpoint common]
\tikzstyle{kpoint conjugate}=[shape=cornerpoint,shorten right=5pt,kpoint common]
\tikzstyle{kpoint transpose}=[shape=cornercopoint,shorten right=5pt,kpoint common]
\tikzstyle{kpoint symm}=[shape=cornerpoint,shorten left=5pt,shorten right=5pt,kpoint common]

\tikzstyle{wide kpoint sc}=[shape=cornerpoint,kpoint common, minimum width=1 cm]
\tikzstyle{wide kpointdag sc}=[shape=cornercopoint,kpoint common, minimum width=1 cm]

\tikzstyle{black kpoint}=[shape=cornerpoint,shorten left=5pt,kpoint common,fill=black,font=\color{white}]

\tikzstyle{black kpoint sm}=[shape=cornerpoint,shorten left=5pt,kpoint common,fill=black,font=\color{white},scale=0.75]

\tikzstyle{black kpoint adjoint}=[shape=cornercopoint,shorten left=5pt,kpoint common,fill=black,font=\color{white}]
\tikzstyle{black kpointadj}=[shape=cornercopoint,shorten left=5pt,kpoint common,fill=black,font=\color{white}]

\tikzstyle{black kpointadj sm}=[shape=cornercopoint,shorten left=5pt,kpoint common,fill=black,font=\color{white},scale=0.75]

\tikzstyle{black dkpoint}=[shape=cornerpoint,shorten left=5pt,kpoint common,fill=black, doubled,font=\color{white}]
\tikzstyle{black dkpoint adjoint}=[shape=cornercopoint,shorten left=5pt,kpoint common,fill=black, doubled,font=\color{white}]
\tikzstyle{black dkpointadj}=[shape=cornercopoint,shorten left=5pt,kpoint common,fill=black, doubled,font=\color{white}]

\tikzstyle{black dkpoint sm}=[shape=cornerpoint,shorten left=5pt,kpoint common,fill=black, doubled,font=\color{white},scale=0.75]
\tikzstyle{black dkpointadj sm}=[shape=cornercopoint,shorten left=5pt,kpoint common,fill=black, doubled,font=\color{white},scale=0.75]

\tikzstyle{kpointdag}=[kpoint adjoint]
\tikzstyle{kpointadj}=[kpoint adjoint]
\tikzstyle{kpointconj}=[kpoint conjugate]
\tikzstyle{kpointtrans}=[kpoint transpose]

\tikzstyle{big kpoint}=[kpoint, minimum width=1.2 cm, minimum height=8mm, inner sep=4pt, text depth=3mm]

\tikzstyle{wide kpoint}=[kpoint, minimum width=1 cm, inner sep=2pt]
\tikzstyle{wide kpointdag}=[kpointdag, minimum width=1 cm, inner sep=2pt]
\tikzstyle{wide kpointconj}=[kpointconj, minimum width=1 cm, inner sep=2pt]
\tikzstyle{wide kpointtrans}=[kpointtrans, minimum width=1 cm, inner sep=2pt]

\tikzstyle{wider kpoint}=[kpoint, minimum width=1.25 cm, inner sep=2pt]
\tikzstyle{wider kpointdag}=[kpointdag, minimum width=1.25 cm, inner sep=2pt]
\tikzstyle{wider kpointconj}=[kpointconj, minimum width=1.25 cm, inner sep=2pt]
\tikzstyle{wider kpointtrans}=[kpointtrans, minimum width=1.25 cm, inner sep=2pt]

\tikzstyle{gray kpoint}=[kpoint,fill=gray!50!white]
\tikzstyle{gray kpointdag}=[kpointdag,fill=gray!50!white]
\tikzstyle{gray kpointadj}=[kpointadj,fill=gray!50!white]
\tikzstyle{gray kpointconj}=[kpointconj,fill=gray!50!white]
\tikzstyle{gray kpointtrans}=[kpointtrans,fill=gray!50!white]

\tikzstyle{gray dkpoint}=[kpoint,fill=gray!50!white,doubled]
\tikzstyle{gray dkpointdag}=[kpointdag,fill=gray!50!white,doubled]
\tikzstyle{gray dkpointadj}=[kpointadj,fill=gray!50!white,doubled]
\tikzstyle{gray dkpointconj}=[kpointconj,fill=gray!50!white,doubled]
\tikzstyle{gray dkpointtrans}=[kpointtrans,fill=gray!50!white,doubled]

\tikzstyle{white label}=[draw,fill=white,rectangle,inner sep=0.7 mm]
\tikzstyle{gray label}=[draw,fill=gray!50!white,rectangle,inner sep=0.7 mm]
\tikzstyle{black label}=[draw,fill=black,rectangle,inner sep=0.7 mm]

\tikzstyle{dkpoint}=[kpoint,doubled]
\tikzstyle{wide dkpoint}=[wide kpoint,doubled]
\tikzstyle{dkpointdag}=[kpoint adjoint,doubled]
\tikzstyle{wide dkpointdag}=[wide kpointdag,doubled]
\tikzstyle{dkcopoint}=[kpoint adjoint,doubled]
\tikzstyle{dkpointadj}=[kpoint adjoint,doubled]
\tikzstyle{dkpointconj}=[kpoint conjugate,doubled]
\tikzstyle{dkpointtrans}=[kpoint transpose,doubled]

\tikzstyle{kscalar}=[kpoint common, shape=EBox, inner xsep=-1pt, inner ysep=3pt,font=\small]
\tikzstyle{kscalarconj}=[kpoint common, shape=WBox, inner xsep=-1pt, inner ysep=3pt,font=\small]

\tikzstyle{spekpoint}=[kpoint sc,minimum height=5mm,inner sep=3pt]
\tikzstyle{spekcopoint}=[kpoint adjoint sc,minimum height=5mm,inner sep=3pt]

\tikzstyle{dspekpoint}=[spekpoint,doubled]
\tikzstyle{dspekcopoint}=[spekcopoint,doubled]

 \tikzstyle{upground}=[circuit ee IEC,thick,ground,rotate=90,scale=2.5]
 \tikzstyle{downground}=[circuit ee IEC,thick,ground,rotate=-90,scale=2.5]
 \tikzstyle{bigground}=[regular polygon,regular polygon sides=3,draw=gray,scale=0.50,inner sep=-0.5pt,minimum width=10mm,fill=gray]

\tikzstyle{arrs}=[-latex,font=\small,auto]
\tikzstyle{arrow plain}=[arrs]
\tikzstyle{arrow dashed}=[dashed,arrs]
\tikzstyle{arrow bold}=[very thick,arrs]
\tikzstyle{arrow hide}=[draw=white!0,-]
\tikzstyle{arrow reverse}=[latex-]
\tikzstyle{cdnode}=[]

\usepackage{appendix}

\begin{document}
\title{The Characterization of Noncontextuality \\ in the Framework of Generalized Probabilistic Theories}

\author{David Schmid}
\email{dschmid@perimeterinstitute.ca}
\affiliation{Perimeter Institute for Theoretical Physics, 31 Caroline Street North, Waterloo, Ontario Canada N2L 2Y5}
\author{John H. Selby}
\email{john.h.selby@gmail.com}
\affiliation{Perimeter Institute for Theoretical Physics, 31 Caroline Street North, Waterloo, Ontario Canada N2L 2Y5}
\affiliation{ICTQT, University of Gda\'nsk, Wita Stwosza 63, 80-308 Gda\'nsk, Poland}
\author{Elie Wolfe}
\affiliation{Perimeter Institute for Theoretical Physics, 31 Caroline Street North, Waterloo, Ontario Canada N2L 2Y5}
\author{Ravi Kunjwal}
\affiliation{Perimeter Institute for Theoretical Physics, 31 Caroline Street North, Waterloo, Ontario Canada N2L 2Y5}
\affiliation{Centre for Quantum Information and Communication, Ecole polytechnique de Bruxelles,
CP 165, Universit\'e libre de Bruxelles, 1050 Brussels, Belgium}
\author{Robert W. Spekkens}
\affiliation{Perimeter Institute for Theoretical Physics, 31 Caroline Street North, Waterloo, Ontario Canada N2L 2Y5}

\date{\today}

\begin{abstract}
 To make precise the sense in which the operational predictions of quantum theory conflict with a classical worldview, it is necessary to articulate a notion of classicality within an operational framework.
 A widely applicable notion of classicality of this sort
  is whether or not the predictions of a given operational theory
   can be explained by a generalized-noncontextual ontological model.
  We here explore what notion of classicality this implies for the generalized probabilistic theory (GPT) that arises from a given operational theory, focusing on prepare-measure scenarios. We first show that, when mapping an operational theory to a GPT by quotienting relative to operational equivalences, the constraint of explainability by a generalized-noncontextual ontological model is mapped to the constraint of explainability by an ontological model. We then show that, under the additional assumption that the ontic state space is of finite cardinality, this constraint on the GPT can be expressed as a geometric condition which we term {\em simplex-embeddability}. Whereas the traditional notion of classicality for a GPT is that its state space be a simplex and its effect space be the dual of this simplex, simplex-embeddability merely requires that its state space be {\em embeddable} in a simplex and its effect space in the dual of that simplex. We argue that simplex-embeddability constitutes an intuitive and freestanding notion of classicality for GPTs.
Our result also has applications to witnessing nonclassicality in prepare-measure experiments.
\end{abstract}
\maketitle

In what precise sense does quantum theory necessitate a departure from a classical worldview? Although this is one of the central questions in the foundations of quantum theory, there is no consensus on its answer. Arguably the two most stringent notions of nonclassicality proposed to date are: the failure to admit of a locally causal ontological model (Bell's theorem)~\cite{Bell,Bellreview} and the failure to admit of a generalized-noncontextual ontological model~\cite{gencontext}.  Both of these are operationally meaningful notions of nonclassicality,  in the sense that one can determine in principle whether a given set of operational statistics admits of a classical explanation by their lights, regardless of its consistency with quantum theory.\footnote{ Refs.~\cite{cabello1998proposed, cabello2000kochen, simon2000feasible, simon2001hidden, larsson2002kochen, cabello2008proposed} provide alternative proposals for how to operationalize the Kochen-Specker notion of noncontextuality~\cite{KS}.
   See Refs.~\cite{operationalks, Mazurek2016, Kunjwal16} for a critique of these.
   } This implies that any experimental evidence for such nonclassicality imposes a constraint on any physical theory that hopes to be empirically adequate, including any putative successor to quantum theory.

For prepare-measure experiments on a single system, the notion of local causality is not applicable, and so, of the two notions, only generalized noncontextuality is a candidate for an operationally meaningful notion of classicality for such experiments.  Elsewhere~\cite{Pirsa_nc} it has been argued that its operational meaningfulness and its larger scope of applicability make the notion of generalized noncontextuality the best notion of classicality available today.  Furthermore, it can be shown to subsume the central ideas behind several other notions of classicality, such as the existence of a nonnegative quasiprobability representation~\cite{negativity,Ferrie_2008}, or of a locally causal model~\cite{Bell,Bellreview}.
 Additionally, the {\em failure} of generalized noncontextuality has been shown to be behind certain notions of nonclassicality, such as anomalous weak values~\cite{AWV, KLP19} and advantages for information processing~\cite{POM,RAC,RAC2,Saha_2019,MESD,cloningcontext,magic,comp1,comp2, YK20}.  

 
Within the framework of ontological models \cite{gencontext}, assuming generalized noncontextuality can be understood as assuming a version of a methodological principle for theory construction due to Leibniz: the ontological identity of empirical indiscernibles (see Ref.~\cite{Leibniz} and the appendix of Ref.~\cite{Mazurek2016}). Given that Einstein made significant use of this principle when he developed the theory of Relativity~\cite{Leibniz}, it is seen to have impressive credentials in physics and therefore is a natural constraint to impose on ontological models.  From this perspective, the impossibility of finding generalized-noncontextual ontological models is best understood as a failing of the framework of ontological models itself, and hence as a type of nonclassicality.

 An operational theory provides an
account of the experimental procedures accessible in the lab and the operational statistics they yield.  A generalized probabilistic theory~\cite{Hardy,Barrett2007} (GPT) is obtained from an operational theory by discarding information about experimental procedures
that can be varied without affecting the operational statistics.  An ontological model of an operational theory or of a GPT is an attempt to provide a realist underpinning to these, that is, a causal account of the statistics they predict.
We will here characterize what the existence of a generalized-noncontextual ontological model of an operational theory implies about the geometry of the GPT associated to the operational theory. Ultimately, we will prove the following:
\begin{theoremNew} \label{thm:Simplex-NC1}
 For a prepare-measure experiment, the operational theory describing it admits of a generalized-noncontextual ontological model on an ontic state space of finite cardinality if and only if the GPT describing it is {\em simplex-embeddable}.
\end{theoremNew}
 Simplex embeddability, defined rigorously in Definition~\ref{embeddabilitydefn}, stipulates that there is a linear map that embeds the state space in a simplex and another linear map that embeds the effect space in the dual of this simplex, such that the pair of maps together preserve inner products.

Hence, if one takes explainability by a generalized-noncontextual model as one's notion of classicality for an operational theory, then one must take simplex-embeddability as one's notion of classicality for a GPT.
This is in contrast with the prevailing idea (see, e.g., \cite{Barrett2007}) that a GPT should be deemed classical if and only if its state space is a simplex and its effect space is the dual thereof,  a condition which we term {\em simpliciality}.  In our conclusions, we discuss the significance of the difference between simplex embeddability and simpliciality as notions of classicality for a GPT. 


We begin by introducing the requisite preliminary concepts.

\noindent{\bf Operational Theories---}
An {\em operational theory} is a minimal type of theory that stipulates, for a given system, a set of preparation procedures and measurement procedures that can be implemented on that system, denoted \textsf{Preps} and \textsf{Mmts} respectively. These are conceptualized as lists of lab instructions that one could implement on the given system. We will here find it useful to consider {\em operational effects}, defined as the tuple consisting of a measurement and an outcome thereof. We obtain the set of all operational effects, denoted \textsf{Effects},  by considering the set of all outcomes $k$ for each measurement $M$ in the set  \textsf{Mmts}.  A particular operational effect will be denoted $[k|M]$.  An operational theory stipulates a probability rule that determines the probability of obtaining operational effect $[k|M]$ given preparation $P$, denoted ${\rm Pr}([k|M],P)$. This probability rule must be compatible with certain relations that hold between the procedures.\footnote{
In previous work (and here), these relations are {\em only} stipulated in  the ``ordinary language'' description of the procedures. In forthcoming work \cite{causalinferential}, we show how they can be incorporated into the formal structure of an operational theory.
}
For instance, if $P_1$ is described as a procedure that convexly mixes $P_2$ and $P_3$, with the choice determined by a coin-flipping mechanism, then ${\rm Pr}([k|M],P_1)$ must be equal to the corresponding mixture of ${\rm Pr}([k|M],P_2)$ and ${\rm Pr}([k|M],P_3)$ for all $[k|M]$~\cite{Hardy}.
For operational effects, analogous constraints from convexity hold, as do additional constraints due to coarse-graining relationships among operational effects. For example, if one operational effect $[k_1|M_1]$ is described as being the coarse-graining of two others, $[k_2|M_2]$ and $[k_3|M_3]$, then ${\rm Pr}([k_1|M_1],P)$ must be the sum of ${\rm Pr}([k_2|M_2],P)$ and ${\rm Pr}([k_3|M_3],P)$ for all $P$.
As we comment below, these constraints on ${\rm Pr}(\_, \_)$ have important consequences for the generalized probabilistic theory associated to the operational theory.

In all, an operational theory of a prepare-measure experiment on a single system is a triple $T:=\{\textsf{Preps}, \textsf{Mmts}, {\rm Pr}(\_, \_)\}$  satisfying these constraints. 

Finally, we define the notion of {\em operational equivalence} of procedures~\cite{gencontext}. Preparation procedures $P$ and $P'$ are said to be operationally equivalent, denoted $P \simeq P'$, if they give rise to the same statistics for all physically possible operational effects, that is, if
${\rm Pr}([k|M],P)={\rm Pr}([k|M],P')$ for all $[k|M]\in \textsf{Effects}$. Operational effects $[k|M]$ and $[k'|M']$ are said to be operationally equivalent, denoted $[k|M] \simeq [k'|M']$, if they give rise to the same statistics for all physically possible preparations, that is, if ${\rm Pr}([k|M],P)={\rm Pr}([k'|M'],P)$ for all $P \in \textsf{Preps}$.

\noindent{\bf Generalized Probabilistic Theories---}
The framework of generalized probabilistic theories  provides a means of describing   the landscape of possible theories of the world, as characterized (solely) by  the operational statistics they predict~\cite{Hardy,Barrett2007}.  
 Quantum and classical theories are included as special cases, but the framework also accommodates alternatives to these.
Although the framework allows for sequential and parallel composition of processes, we will focus on the fragment of a GPT that describes prepare-measure experiments on a single system. 

 A given GPT associates to a system a convex set of {\em states},  $\Omega$. One can think of this set as being a generalization of the Bloch ball in quantum theory, where the states in the set are the normalised (potentially mixed) states of the theory. We make the standard assumptions that $\Omega$ is finite dimensional and compact. While $\Omega$ naturally lives inside an affine space, $\mathsf{AffSpan}[\Omega]$, for convenience we will represent it as living inside a real inner product space $(V, \left<\_,\_\right>)$ of one dimension higher, where we embed $\mathsf{AffSpan}[\Omega]$ as a hyperplane in $V$ which does not intersect with the origin $\boldsymbol{0}$. This is analogous to embedding the Bloch-Ball within the real vector space of Hermitian matrices. The reason for doing so is that we can then define both the GPT states and GPT effects within the same space.
 
 Note that we will {\em not} restrict attention to GPTs satisfying the no-restriction hypothesis~\cite{chiribella2010probabilistic}, which stipulates that all the states and effects that are logically possible must also be physically possible.

A GPT also associates to every system a set of GPT {\em effect vectors}, $\mathcal{E}$.  In the framework of GPTs, the probability of obtaining an effect $\boldsymbol{e}\in\mathcal{E}$
given a state $\boldsymbol{s}\in\Omega$ is given by the inner product:
\beq
\mathrm{Prob}(\boldsymbol{e},\boldsymbol{s}) := \left< \boldsymbol{e},\boldsymbol{s}\right>.
\eeq
 We require that $\mathcal{E}$ must satisfy the following constraints.
If one defines the dual of $\Omega$, denoted $\Omega^*$,  as the set of vectors in $V$ whose inner product with all state vectors in $\GPTt$ is between $0$ and $1$, i.e.,
\beq
\Omega^* := \{ \boldsymbol{x}\in V | \left<\boldsymbol{x},\boldsymbol{s}\right> \in [0,1] \ \forall \boldsymbol{s} \in \Omega\},
\eeq
then $\mathcal{E}$ is a compact convex set contained in $\Omega^*$, $\mathcal{E} \subseteq \Omega^*$,
which  contains the origin $\boldsymbol{0}$ and the ``unit effect'' $\boldsymbol{u}$, which in turn satisfy, respectively, $\left<\boldsymbol{0},\boldsymbol{s}\right> = 0$ and $\left<\boldsymbol{u},\boldsymbol{s}\right> = 1$ for all $\boldsymbol{s}\in\Omega$.
Due to how we embedded $\mathsf{AffSpan}[\Omega]$ within $V$, $\boldsymbol{u}$ necessarily exists and is unique.\footnote{This uniqueness is referred to as the ``causality axiom'' in Ref.~\cite{chiribella2010probabilistic}.}

The state and effect spaces of any valid GPT must satisfy the principle of {\em tomography}, which states that the GPT states and GPT effects can be uniquely identified by the probabilities that they produce. Formally, for the GPT states, we have  that $\left<\gp{e},\gp{s_1}\right> = \left<\gp{e},\gp{s_2}\right>$ for all $\gp{e}\in\gpEff$ if and only if $\gp{s_1} = \gp{s_2}$, and for the GPT effects, we have that $\left<\gp{e_1},\gp{s}\right> = \left<\gp{e_2},\gp{s}\right>$ for all $\gp{s}\in\GPTt$ if and only if $\gp{e_1} = \gp{e_2}$.

A GPT $G$, therefore, is defined as a quadruple $G:=(V, \left<\_,\_\right>, \Omega, \mathcal{E})$ satisfying these constraints.

\noindent{\bf The GPT associated to an Operational Theory---}
The GPT
associated to an operational theory $T$ is the theory that one obtains when one quotients $T$ relative to the notion of operational equivalence defined above.
It is specified by a pair of quotienting maps
\beq
\boldsymbol{s}_{\_}: \textsf{Preps} \to \Omega \quad\text{and}\quad \boldsymbol{e}_{\_}: \textsf{Effects} \to \mathcal{E}
\eeq
taking each preparation $P$ (operational effect $[k|M]$) to a GPT state vector $\boldsymbol{s}_{P}$ (GPT effect vector $\boldsymbol{e}_{[k|M]}$) representing its operational equivalence class.
The maps jointly satisfy the constraint that
\beq \label{predreprod}
{\rm Pr}([k|M],P) =  \left< \boldsymbol{e}_{[k|M]}, \boldsymbol{s}_{P}\right>
\eeq
for all preparations and effects in the operational theory.

Note that Eq.~\eqref{predreprod} and the assumption of tomography guarantee that every operationally equivalent pair of preparations (effects) in the operational theory is mapped to the same GPT state (GPT effect) vector, and hence that each GPT vector is a representation of an operational equivalence class of operational procedures. That is:
\begin{align}
P &\simeq P' & &\iff & \boldsymbol{s}_{P}&=\boldsymbol{s}_{P'}\ \text{and}\\
[k|M]&\simeq [k'|M'] & &\iff & \boldsymbol{e}_{[k|M]}&=\boldsymbol{e}_{[k'|M']}
\end{align}
Furthermore, they imply that nontrivial convex and coarse-graining relations holding among preparations (respectively effects) in the operational theory are encoded in the geometric relations between the GPT state vectors (respectively GPT effect vectors) in the GPT.
For example, if $P_1$ is a convex mixture of $P_2$ and $P_3$ with weights $w$ and $(1-w)$ (or if $P_1$ is operationally equivalent to such a mixture), then it will follow that $\boldsymbol{s}_{P_1} = w \boldsymbol{s}_{P_2} +(1-w) \boldsymbol{s}_{P_3}$.

\noindent{\bf Ontological Model of an Operational Theory---}

  An {\em ontological model of an operational theory $T$} is an attempt to  provide a causal explanation of
 the operational statistics of $T$.  For a prepare-measure experiment, it posits that
 the response of the measurement is determined (possibly probabilistically) by the ontic state $\lambda$ of the system (a complete characterization of its physical attributes), while preparation procedures determine the distribution over the space of ontic states, $\Lambda$, from which $\lambda$ is sampled.

%

 More precisely,   an ontological model associates to each preparation $P\in \textsf{Preps}$ a normalized probability distribution over $\Lambda$, denoted $\mu_{P}$, representing an agent's knowledge of the ontic state when they know that the preparation was $P$. Denoting the set of such distributions by $\mathcal{D}[\Lambda]$, the ontological model specifies a map
\beq\label{OMmu}
\mu_{\_}: \textsf{Preps} \to \mathcal{D}[\Lambda].
\eeq
Furthermore, an ontological model associates to each operational effect $[k|M] \in \textsf{Effects}$ a response function on $\Lambda$, denoted $\xi_{[k|M]}$, where $\xi_{[k|M]}(\lambda)$ represents the probability  assigned to  
  the outcome $k$ in a measurement of $M$
 if the ontic state of the system fed into the measurement device
   were known to be   $\lambda\in \Lambda$.  Denoting the set of such response functions by $\mathcal{F}[\Lambda]$, the ontological model specifies a map
 \beq\label{OMxi}
 \xi_{\_}: \textsf{Effects} \to \mathcal{F}[\Lambda].
 \eeq
These two maps must preserve the convex and coarse-graining relations between operational procedures that were discussed above.
For example, if $P_1$ is a convex mixture of $P_2$ and $P_3$ with weights $w$ and $(1-w)$, then $\mu_{P_1} = w \mu_{P_2} +(1-w) \mu_{P_3}$~\cite{gencontext}, and similarly for operational effects. Finally, the ontological model must reproduce the probability rule of the operational theory $T$ via
\begin{equation}\label{OMprob}
{\rm Pr}([k|M],P)= \sum_{\lambda \in \Lambda} \xi_{[k|M]}(\lambda) \mu_P(\lambda).
\end{equation}
(where we have assumed $\Lambda$ to be discrete for simplicity).

\noindent{\bf Generalized Noncontextuality---} We are now in a position to define the notion of classicality of an operational theory $T$ with which we are concerned in this article, namely, the existence of a generalized-noncontextual ontological model of $T$.  An ontological model of a prepare-measure experiment satisfies generalized noncontextuality if every two procedures which are operationally equivalent have identical representations in the ontological model.

In other words, the constraint for preparations is that
\beq
P \simeq P' \ \ \implies \ \ \mu_{P} =\mu_{P'},
\eeq
while the constraint for operational effects is that
\beq
[k|M] \simeq [k'|M'] \ \ \implies \ \ \xi_{[k|M]} =\xi_{[k'|M']}.
\eeq
These constraints formalize the Leibnizian principle discussed in the introduction insofar as the empirical indiscernibility of procedures (the antecedents) imply the equality of their ontological representations (the consequents).  

\noindent{\bf Ontological Model of a GPT---} 
 As is the case for   an ontological model of an operational theory, an ontological model of a GPT is an attempt  to provide a causal explanation of   the operational statistics in terms of a space $\Lambda$ of ontic states  for the system.  
In this case, however, what is being modelled ontologically are not preparations and measurements, but operational equivalence classes thereof.
Thus, an ontological model of a GPT associates to each GPT state vector $\boldsymbol{s} \in \Omega$ a normalized probability distribution over $\Lambda$, denoted $\tilde{\mu}_{\boldsymbol{s}}\in\mathcal{D}[\Lambda]$, and to each GPT effect vector $\boldsymbol{e} \in \mathcal{E}$ a response function on $\Lambda$, denoted $\tilde{\xi}_{\boldsymbol{e}}\in\mathcal{F}[\Lambda]$.
Hence, it specifies a pair of maps
\beq\label{OMGPTmu}
\tilde{\mu}_{\_}: \Omega \to \mathcal{D}[\Lambda]\quad \text{and}\quad \tilde{\xi}_{\_}: \mathcal{E}  \to \mathcal{F}[\Lambda],
\eeq
which must be linear by the assumption that they preserve the convex and coarse-graining relations defined by the geometry of the GPT state and GPT effect spaces.
Finally, the ontological model must reproduce the probability rule of the GPT via
\begin{equation}\label{OMGPTprob}
\left< \boldsymbol{e},\boldsymbol{s} \right> = \sum_{\lambda \in \Lambda} \tilde{\xi}_{\boldsymbol{e}}(\lambda) \tilde{\mu}_{\boldsymbol{s}}(\lambda).
\end{equation}

It is now clear that a generalized-noncontextual ontological model of an operational theory $T$ is equivalent to an ontological model of the GPT associated to $T$. (An explicit proof is given in the Supplemental Material.) Hence:
\begin{proposition}\label{Bridge}
There exists a generalized-noncontextual ontological model of an operational theory $T$ describing prepare-measure experiments on a system if and only if there exists an ontological model of the GPT $G$ that $T$ defines.
\end{proposition}

As we show in the Supplementary Material,
an ontological model of a GPT is equivalent to a positive quasiprobability representation of that GPT. Hence,
Proposition~\ref{Bridge} is the generalization of the results of Refs.~\cite{negativity,Ferrie_2008} from quantum theory to an arbitrary GPT.

Note that ontological models of GPTs, unlike those of operational theories, cannot be said to be either generalized-contextual or generalized-noncontextual.
Recall that contexts are defined as differences among procedures that are operationally equivalent, so there is no notion of context in a GPT, since the latter is obtained by quotienting relative to operational equivalences.
To ask whether the ontological representation of a GPT state (or GPT effect) varies with context is a category mistake since there is no variability of context for GPT states (or GPT effects), just as 
it is a category mistake to ask whether $X$ varies with   $Y$ when $Y$ exhibits no variability~\cite{sep-category-mistakes}.

This implies another contrast between ontological models of GPTs and those of operational theories.
For a given operational theory, one can {\em always} construct an ontological model by allowing this model to be generalized-contextual (using the analogue of the construction of   Ref.~\cite{Beltrametti_1995}).  But it is {\em not} the case that one can always construct an ontological model of a GPT because such models do not have the benefit of the representational flexibility afforded by nontrivial context-dependences. Indeed, it is this lack of flexibility that implies the necessity of negativity in quasiprobability representations of some GPTs~\cite{negativity,Ferrie_2008}.

\noindent{\bf The Geometric Criterion associated to Noncontextuality---}
We argued in the introduction that an operational theory is best viewed as classical if its operational predictions admit of an explanation in terms of a generalized-noncontextual ontological model.
It is natural, therefore, to determine what this notion of classicality for an operational theory entails for the GPT that the latter defines.  By Proposition~\ref{Bridge}, this is equivalent to finding a criterion for when a GPT admits of an ontological model.
We now give such a condition, which we term {\em simplex-embeddability},  under the assumption that the ontological model has an ontic state space $\Lambda$ of finite cardinality.
  At the end of the article, we mention  some follow-up  
works that lift this restriction. 

\begin{definitionNew}[Simplex-embeddable GPTs] \label{embeddabilitydefn}
      A GPT describing a prepare-measure experiment, $G=(V,\left<\_,\_\right>_V,\Omega, \mathcal{E})$, is {\em simplex-embeddable} iff there exists (i) an inner product space $(W,\left<\_,\_\right>_W)$ of some dimension $d$ which contains a $(d-1)$-dimensional (hence $d$-vertex) simplex $\Delta_{d}$ (whose affine span does not contain the origin) and its dual hypercube
      $\Delta_{d}^*$, and
      (ii) a pair of linear maps $\iota,\kappa : V\to W$ satisfying
        \begin{align}
        \iota(\Omega) &\subseteq  \Delta_{d}, \\
        \kappa(\mathcal{E})&\subseteq \Delta_{d}^*, \text{ and} \\
        \left<{\boldsymbol{e}},{\boldsymbol{s}}\right>_V &= \left<\kappa({\boldsymbol{e}}),\iota({\boldsymbol{s}})\right>_W \ \ \forall {\boldsymbol{e}}\in\mathcal{E}\ , \ \forall {\boldsymbol{s}}\in\Omega.
        \end{align}
    \end{definitionNew}
\noindent Note that while it is only the space of GPT {\em states} which embeds within a simplex, while the space of GPT effects embeds within a hypercube dual to this simplex,
we nonetheless use the term `simplex-embeddable'  as an umbrella term for the pair of embedding relations.

With this definition in hand, one can prove the following (as we show in the Supplementary Material).
\begin{theoremNew}
\label{thm:Simplex-NC}
A GPT describing a prepare-measure experiment admits of an ontological model over an ontic space $\Lambda$ of finite cardinality if and only if it is simplex-embeddable.
\end{theoremNew}
Crucially, note that the dimension of the vector space in which this embedding can be constructed may be greater than the native dimension of the GPT. We provide an explicit example of the necessity of such a `dimension gap' in the Supplementary Material, Section~\ref{app:dimensionmismatch}.

By combining Proposition~\ref{Bridge} and Theorem~\ref{thm:Simplex-NC}, one immediately obtains our main result, Theorem~\ref{thm:Simplex-NC1}, which gives a geometric characterization of the set of GPTs that are associated to operational theories which admit of generalized-noncontextual ontological models over ontic state spaces of finite cardinality.

\noindent{\bf Discussion---}
 As noted earlier, the prevailing view up to now has been  that a GPT should be deemed classical if and only if it is simplicial, i.e., if its state space is a simplex and its effect space is the hypercube that is dual to this simplex.
To understand the distinction between simpliciality and simplex-embeddability, we recall
a distinction between notions of nonclassicality introduced in Ref.~\cite{epistricted}\footnote{ Note, however, that there it was the nonclassicality of operational phenomena rather than of operational theories that was at issue.}:
an operational theory is deemed {\em weakly nonclassical} if it exhibits measurement incompatibility (sets of measurements that cannot all be simulated by processing the outcome of a single measurement) or ambiguity of mixtures (mixed states with multiple convex decompositions into pure states) or both,
while it is deemed {\em strongly nonclassical} if it furthermore fails to admit of a generalized-noncontextual ontological model.
This distinction can be extended from operational theories to GPTs using Proposition~\ref{Bridge}:
a GPT is deemed {\em weakly nonclassical} if it exhibits measurement incompatibility\footnote{In our formalization of GPTs, a measurement is any collection of effects that sum to the unit effect.} or ambiguity of mixtures or both, while it is deemed {\em strongly nonclassical} if it furthermore fails to admit of an ontological model.
Because a GPT exhibits measurement incompatibility if and only if its effect space is not a hypercube~\cite{plavala2016all} and it exhibits ambiguity of mixtures if and only if its state space is not a simplex,
   it follows that
the notion of nonclassicality captured by nonsimpliciality of a GPT (the focus of previous work) is merely weak nonclassicality.
By contrast, the notion of nonclassicality captured by the failure of {\em simplex embeddability} (introduced here) is exactly that of strong nonclassicality.

For both operational theories and GPTs, the strong notion of nonclassicality captures the idea that operational predictions {\em resist a classical explanation}.
 In light of Theorem~\ref{thm:Simplex-NC1}, each of the motivations listed in the introduction for taking generalized noncontextuality as a good notion of classical explainability {\em for operational theories} can be reappropriated as a motivation for taking simplex-embeddability as a good notion of classical explainability {\em for a GPT}.
In the reverse direction, Theorem~\ref{thm:Simplex-NC1} provides a novel and independently motivated justification for generalized noncontextuality as a good notion of classical explainability of an operational theory, since at the level of the GPT, simplex-embeddability is a very natural way to formalize the notion of classical explainability.

   In related work undertaken simultaneously~\cite{Shahandeh}, Shahandeh considers the consequences of a notion of noncontextuality, termed {\em broad noncontextuality}, which differs from the notion of generalized noncontextuality.   As we show in the Supplementary Material, our results imply that the geometric condition on a GPT associated to this notion is simplex embeddability {\em without a dimension gap}.
Ref.~\cite{Shahandeh} focuses on GPTs that satisfy the no-restriction hypothesis~\cite{chiribella2010probabilistic}  
and demonstrates that for such GPTs, simplex-embeddability without a dimension gap coincides with simpliciality.\footnote{For many years, the no-restriction hypothesis was taken to be a fundamental part of the definition of a GPT (indeed, it was only in Ref.~\cite{chiribella2010probabilistic} in 2009 that the term `no-restriction hypothesis' was coined), but it was later acknowledged to be a substantial restriction on the sorts of GPTs which could be considered.   Examples of GPTs that violate the no-restriction hypothesis include: the GPT associated to Spekkens' toy model \cite{spekkens2007evidence}, the stabilizer subtheories of quantum theory as well as the Gaussian and quadrature subtheories of quantum mechanics~\cite{epistricted}, 
  and the GPTs considered in, for example, Refs.\cite{janotta2013generalized,janotta2014non,mazurek2017experimentally,sainz2018almost,filippov2020operational,wright2020general}.}
We discuss the relation between Ref.~\cite{Shahandeh} and this work further in the Supplementary Material, and also demonstrate that it is {\em only} for GPTs satisfying the no-restriction hypothesis that simplex embeddability and simpliciality coincide.  Barnum and Lami~\cite{BarnumLami} have found
  similar results to those presented here and  in Ref.~\cite{Shahandeh}, while also considering generalizations to infinite dimensions.

\noindent{\bf Applications---} Our result provides a novel way to 
 test, for   a given set of 
 experimentally realized preparations and measurements which are tomographically complete,  whether these provide evidence of strong nonclassicality or not.   
One first determines the  set of GPTs that are compatible with the data obtained from prepare-measure experiments,   using the techniques described in Ref.~\cite{mazurek2017experimentally}  (because of noise and finite precision effects, it is never a {\em single} GPT that is picked out by the data). 
One then tests these GPTs for simplex-embeddability.

 The previous gold standard for testing generalized noncontextuality~\cite{Mazurek2016,PuseydelRio} was to test a specific noise-robust noncontextuality inequality, obtained from a specific set of operational equivalences\footnote{This set is usually inferred from a pre-existing no-go theorem for generalized noncontextuality in quantum theory, for instance, those inspired by Kochen-Specker contradictions \cite{operationalks, stated, Kunjwal19, Kunjwal20}.}. Such a test, however, requires an experimentalist to target a set of preparations and a set of measurements satisfying these specific operational equivalences.\footnote{In fact, because the procedures that are experimentally realized always deviate from the targets, to satisfy the operational equivalences, one must shift attention to ``secondary'' procedures which lie within the convex hull of the realized ones, a step that compromises some purity~\cite{robust}.} 
 The simplex-embedding test for generalized noncontextuality, by contrast, makes use of {\em whatever} operational equivalences happen to be satisfied by the preparations and measurements that were physically realized.  It also makes use of {\em all} such equivalences, since these relations are encoded in the geometry of the state and effect spaces.   Consequently, one needn't design an experiment to target particular equivalences.\footnote{Nor post-process the data to satisfy them.}
Rather, the technique can be applied to data obtained in {\em any} experiment that achieves tomographic completeness, including those not dedicated to testing noncontextuality.  The technique's scope of applicability is therefore much greater and consequently promises greater applications. For instance, our method can be used to determine whether prepare-measure data obtained in some experimental architecture for quantum computation (e.g., in a noisy implementation of some simple quantum algorithm) either does or does not witness the presence of strong nonclassicality.

 It is worth noting two practical issues with testing for simplex-embeddability. Firstly, one needs some  algorithm for testing 
 whether a generic GPT $G$  (described geometrically) 
 admits an embedding of the type described in Definition~\ref{embeddabilitydefn} for a given dimension $d$. Secondly, even given such an algorithm, the question arises of whether there is an  \emph{upper bound} on the dimension $d$ up to which one must apply this test. The latter question has been addressed by subsequent work, described below.

Note that to merely witness nonclassicality of a given
 GPT, it suffices to find inner approximations of its state and effect spaces which do  not embed in a simplex and its dual. Similarly, to merely witness classicality,
 it suffices to find outer approximations of these that {\em do} embed
  in a simplex and its dual.
The problem of finding such witnesses can be
 simplified, therefore, by making a propitious choice of the shape of these inner or outer approximations.
This trick can also be applied to experimental data---
it suffices to choose convenient inner and outer approximations to the spaces 
of {\em all} the GPTs in the set compatible with the data.

\noindent{\bf Subsequent work---}
Two of the key questions raised by this work have been resolved by subsequent works. First, the results of Refs.~\cite{Woods2020} and ~\cite{Schmid2020} imply that the dimensional caveats in our main theorems can be dropped, removing the primary obstacle to testing for simplex-embeddability in practice. In particular, Corollary~33 of Ref.~\cite{Woods2020} provides a dimension bound given by the square of the GPT's dimension. For tomographically local~\cite{hardy2011reformulating} GPTs, Ref.~\cite{Schmid2020} provides a tight bound given by the GPT's dimension itself (under some natural assumptions).   
(It follows that every GPT of finite dimension that admits of an ontological model admits of one with a finite number of ontic states.)
  Second, Ref.~\cite{Schmid2020} extends our result beyond the prepare-measure scenario to general compositional scenarios.

\begin{acknowledgments}
We thank Matt Pusey and Thomas Galley for helpful discussions. RWS thanks Howard Barnum, Alex Wilce, and Jonathan Barrett for early discussions on the topic of this article.  D.S. is supported by a Vanier Canada Graduate Scholarship. JHS was supported in part by the Foundation for Polish Science through IRAP project co-financed by EU within Smart Growth Operational Programme (contract no. 2018/MAB/5). RK is supported by the Program of Concerted Research Actions (ARC) of the Universit\'e libre de Bruxelles. This research was supported by Perimeter Institute for Theoretical Physics. Research at Perimeter Institute is supported in part by the Government of Canada through the Department of Innovation, Science and Economic Development Canada and by the Province of Ontario through the Ministry of Colleges and Universities.
\end{acknowledgments}

\bigskip
\setlength{\bibsep}{1pt plus 1pt minus 1pt}
\bibliographystyle{apsrev4-2-wolfe}
\nocite{apsrev41Control}
\bibliography{bibliography}

\appendix
\appendixpage

\section{Proof of Proposition~\ref{Bridge} }

We now prove Proposition~\ref{Bridge}.
\proof
 In one direction, given an ontological model of the GPT $G$ which is associated to the operational theory $T$ (via the quotienting maps $\boldsymbol{s}_{\_}$ and $\boldsymbol{e}_{\_}$), i.e., given the relevant maps $\tilde{\mu}_{\_}$ and $\tilde{\xi}_{\_}$,  we can construct a generalized-noncontextual ontological model for $T$ by simply composing the quotienting map followed by the ontological map; that is, by constructing
\beq
\mu_{\_} :=\tilde{\mu}_{ \boldsymbol{s}_{\_}}
\eeq
and
\beq
 \xi_{\_}:=\tilde{\xi}_{\boldsymbol{e}_{\_}}.
 \eeq
 It is then easy to check that these maps define an ontological model of the operational theory $T$. Firstly, $\mu_{\_}$ preserves the convex relations among preparations, since $\boldsymbol{s}_{\_}$ preserves these relations and $\tilde{\mu}_{\_}$ is linear. Similarly, $\xi_{\_}$ preserves the convex and coarse-graining relations for operational effects.
The resulting model is easily seen to be noncontextual, since
\begin{align}
P_1\simeq P_2 &\implies \boldsymbol{s}_{P_1} = \boldsymbol{s}_{P_2}\\ &\implies \tilde{\mu}_{\boldsymbol{s}_{P_1}} = \tilde{\mu}_{\boldsymbol{s}_{P_2}}\\ &\implies \mu_{P_1} = \mu_{P_2},
\end{align}
and similarly for operational effects.
Finally, the two maps together reproduce the predictions of the operational theory, since
\begin{align}
\sum_{\lambda \in \Lambda} \xi_{[k|M]}(\lambda) \mu_P(\lambda) &= \sum_{\lambda \in \Lambda} \tilde{\xi}_{\boldsymbol{e}_{[k|M]}}(\lambda) \tilde{\mu}_{ \boldsymbol{s}_{P}}(\lambda) \\ &\stackrel{\ref{OMGPTprob}}{=} \left<\boldsymbol{e}_{[k|M]},\boldsymbol{s}_{P}\right>\\ &\stackrel{\ref{predreprod}}{=} {\rm Pr}([k|M],P).
\end{align}

Conversely, given a generalized-noncontextual ontological model of an operational theory $T$ (i.e., given the relevant maps $\mu_{\_}$ and $\xi_{\_}$), we can construct an ontological model of the GPT $G$ associated to it (via the relevant maps $\boldsymbol{s}_{\_}$ and $\boldsymbol{e}_{\_}$) by defining the maps $\tilde{\mu}_{\_}$ and $\tilde{\xi}_{\_}$ as the unique linear maps satisfying
\beq
\tilde{\mu}_{\boldsymbol{s}} :=\mu_{P}
\eeq
for all preparations $P$ such that $\boldsymbol{s}_{P} =\boldsymbol{s}$, and
\beq
\tilde{\xi}_{\boldsymbol{e}} :=\xi_{[k|M]}
\eeq
for all operational effects $[k|M]$ such that $\boldsymbol{e}_{[k|M]} =\boldsymbol{e}$. It is then easy to check that $\tilde{\mu}_{\_}$ preserves the convex relations between state vectors.
E.g., suppose that $P_1$ is a convex mixture of $P_2$ and $P_3$ with weights $w$ and $(1-w)$, so that $\mu_{P_1} = w \mu_{P_2} +(1-w) \mu_{P_3}$ (since $\mu_{\_}$ preserves the convex relations among preparation procedures). We therefore find that $\tilde{\mu}_{\boldsymbol{s}_{P_1}} = \mu_{P_1} = w \mu_{P_2} +(1-w)\mu_{P_3} = w \tilde{\mu}_{\boldsymbol{s}_{P_2}} +(1-w)\tilde{\mu}_{\boldsymbol{s}_{P_3}}$.
 Thus, in the example under consideration, $\boldsymbol{s}_{P_1} = w \boldsymbol{s}_{P_2} +(1-w) \boldsymbol{s}_{P_3}$ (given the operational equivalence of $P_1$ with the mixture of $P_2$ and $P_3$), and \blk
 $\tilde{\mu}_{\boldsymbol{s}_{P_1}} = w \tilde{\mu}_{\boldsymbol{s}_{P_2}} +(1-w)\tilde{\mu}_{\boldsymbol{s}_{P_3}}$, and so $\tilde{\mu}_{\_}$ does indeed preserve the convex relations among  GPT states.
The proof that $\tilde{\xi}_{\_}$ preserves the convex and coarse-graining relations among GPT effects is analogous.

Finally, the two maps together reproduce the predictions of the GPT, since
\begin{align}
\sum_{\lambda \in \Lambda} \tilde{\xi}_{\boldsymbol{e}}(\lambda)\tilde{\mu}_{\boldsymbol{s}}(\lambda) &= \sum_{\lambda\in\Lambda} \xi_{[k|M]}(\lambda) \mu_{P}(\lambda) \\
&\stackrel{\ref{OMprob}}{=} \mathrm{Pr}(P,[k|M]) \\
&\stackrel{\ref{predreprod}}{=} \left<\boldsymbol{e}_{[k|M]} ,\boldsymbol{s}_{P}\right> \\
&= \left<\boldsymbol{e} ,\boldsymbol{s}\right>.
\end{align}
\endproof

\section{Quasiprobabilistic representations of GPTs} \label{QPrepnsGPTs}

We have argued that a GPT should be deemed classical if and only if an ontological representation of it exists. We now prove the following proposition, which implies that this notion of classicality is equivalent to another notion of classicality, namely, the existence of a positive quasiprobability representation.

 \begin{proposition}\label{OMequivalenttoQP}
An ontological model of a GPT $G$ describing prepare-measure experiments on a system is equivalent to a positive quasiprobability representation thereof. 
 \end{proposition}

A quasiprobability representation of a GPT associates to each
system a set $\Lambda$.
It associates to each GPT state on this system, ${\boldsymbol{s}}\in\Omega$,
a real-valued function over $\Lambda$, denoted $\hat{\mu}_{\boldsymbol{s}}$ and satisfying $\sum_{\lambda \in \Lambda} \hat{\mu}_{\boldsymbol{s}}(\lambda) = 1$.  This is termed a {\em quasiprobability distribution} over $\Lambda$ because if the function were valued in $[0,1]$ rather than the reals, it could be interpreted as a probability distribution over $\Lambda$.  Formally, it specifies a linear map from state vectors to the real vector space of functions from $\Lambda$ to $\mathds{R}$, denoted $\mathds{R}^\Lambda$. That is,
\beq\label{QPGPTmu}
\hat{\mu}_{\_} :\Omega \to\mathds{R}^\Lambda.
\eeq
It also associates to each GPT effect ${\boldsymbol{e}}$ a real-valued function over $\Lambda$, denoted $\hat{\xi}_{\boldsymbol{e}}$ and satisfying
$\sum_{{\boldsymbol{e}} \in \chi} \hat{\xi}_{\boldsymbol{e}}(\lambda) = 1$ for all $\lambda \in \Lambda$ where $\chi$ is any set of GPT effects corresponding to the set of outcomes of a possible measurement. Note that such a set must satisfy $\sum_{{\boldsymbol{e}} \in \chi} {\boldsymbol{e}} = {\bf u}$.
The function $\hat{\xi}_{\boldsymbol{e}}$ is termed a {\em quasiresponse function} because if the function were valued in $[0,1]$ rather than the reals, it could be interpreted as a response function.
  Formally, it specifies a linear map
\beq\label{QPGPTxi}
\hat{\xi}_{\_} :\mathcal{E} \to\mathds{R}^\Lambda.
\eeq
Now consider the probability of obtaining GPT effect ${\boldsymbol{e}}$
given a preparation associated to state ${\boldsymbol{s}}$, which is given by $\left<{\boldsymbol{e}},{\boldsymbol{s}}\right>_V$ in the GPT.  In a quasiprobability representation of this GPT, one computes this probability using the same formula that would be appropriate if the quasi-probabilities were true probabilities, namely, $\sum_{\lambda\in\Lambda}\hat{\xi}_{\boldsymbol{e}}(\lambda)\hat{\mu}_{\boldsymbol{s}}(\lambda)$.  Thus, $\forall {\boldsymbol{e}}\in\mathcal{E}$, $\forall {\boldsymbol{s}}\in\Omega$,
\begin{align}\label{QPGPTprob}
\left<{\boldsymbol{e}},{\boldsymbol{s}}\right>_V &= \hat{\xi}_{\boldsymbol{e}}\cdot \hat{\mu}_{\boldsymbol{s}}=  \sum_{\lambda\in\Lambda}\hat{\xi}_{\boldsymbol{e}}(\lambda)\hat{\mu}_{\boldsymbol{s}}(\lambda).
\end{align}

Finally, we will say that a quasiprobability representation of a GPT is {\em positive}
if $\forall {\boldsymbol{s}}\in \Omega, \forall \lambda\in \Lambda$: $0 \le \hat{\mu}_{\boldsymbol{s}}(\lambda) \le 1$  and
$\forall {\boldsymbol{e}}\in \mathcal{E},\forall \lambda\in \Lambda$: $0 \le \hat{\xi}_{\boldsymbol{e}}(\lambda)\le 1$.

Prop~\ref{OMequivalenttoQP} follows immediately
by noting that Eqs.~\eqref{QPGPTmu},~\eqref{QPGPTxi} together with the positivity constraints are equivalent to Eq.~\eqref{OMGPTmu}, and that
 Eq.~\eqref{QPGPTprob} is equivalent to Eq.~\eqref{OMGPTprob}.

\section{Proof of Theorem~\ref{thm:Simplex-NC}}
We now wish to prove Theorem~\ref{thm:Simplex-NC}.

It is convenient for the proof to formally define the notion of a {\em simplicial GPT} which was discussed in the main text. A simplicial GPT of given dimension $d$ is a tuple  $(W,\left<\_,\_\right>_W, \Delta_{d},\Delta_{d}^*)$ defined by a vector space $W$ with inner product $\left<\_,\_\right>_W$, a simplicial state space $\Omega = \Delta_{d}$ (of intrinsic dimension $d-1$), and the dual hypercube $\mathcal{E}=\Delta_{d}^*$ (of intrinsic dimension $d$) as its effect space. Clearly, the inner product space, simplex, and the dual appearing in the definition of simplex embeddability (Def.~\ref{embeddabilitydefn}) define such a simplicial GPT.\footnote{For a GPT $G$ which is simplex-embeddable but not itself simplicial, the embedding relation cannot be interpreted as merely due to technological limitations.  To posit that a world is governed by the GPT $G$ is to posit that the states and effects that are physically realizable are all and only the states and effects included in $G$.  Consequently, any states and effects that are included in the simplicial GPT into which $G$ embeds, but are outside of $G$, are by assumption {\em not} physically realizable in a world governed by the GPT $G$.
 The embedding relation must therefore be understood as a fundamental (rather than practical) limitation on what is possible in such a world. }

Next, it is useful to characterize the inherent degeneracies in the definition of a GPT, that is, to characterize under what conditions two GPTs make the same operational predictions, and hence should be deemed 
 equivalent.
\begin{definitionNew}[Equivalent GPTs]\label{equivGPT}
Two GPTs, $G=(V,\left<\_,\_\right>_V, \Omega, \mathcal{E})$ and $G'=(V',\left<\_,\_\right>_{V'}, \Omega', \mathcal{E}')$, are said to be equivalent iff there exist linear isomorphisms $\omega, \epsilon : V\to V'$ such that:
\begin{align}
\omega(\Omega) &= \Omega', \\
\epsilon(\mathcal{E})&= \mathcal{E}',\ \text{and} \\
\left<\boldsymbol{e},\boldsymbol{s}\right>_V &= \left<\epsilon_{\boldsymbol{e}},\omega_{\boldsymbol{s}}\right>_{V'} \ \ \forall \boldsymbol{e}\in\mathcal{E}\ , \ \boldsymbol{s}\in\Omega.
\end{align}
\end{definitionNew}
 \noindent As a simple example, note that the GPT $(V,\left<\_,\_\right>_V, \Omega, T(\mathcal{E}))$, where $T$ is a reversible map, is equivalent to the GPT  $(V,\left<\_,\_\right>_V, T^{\dag}(\Omega), \mathcal{E})$ where $\dag$ denotes the adjoint operation relative to the inner product on the vector space  
   ~\cite{Hardy}.

Now, note that any simplicial GPT of a given dimension $d$ is equivalent (in the above sense) to one in a particular canonical form, defined as follows.  The vector space is taken to be $\mathds{R}^d$ with some chosen orthonormal basis $\{\boldsymbol{b}_i\}_{i\in B}$ and the inner product is taken to be the dot-product with respect to this basis. The state space is the unit simplex (with $d$ vertices and intrinsic dimension $d-1$), defined as
\beq
\Delta_{d} := \mathsf{ConvexHull}[\{\boldsymbol{b}_i\}],
\eeq
and the effect space is the dual of it, namely the unit hypercube (with $2^{d-1}$ vertices and intrinsic dimension $d$)
\beq
\Delta_{d}^* = \mathsf{ConvexHull}[\{\textstyle\sum_{i\in \chi}\boldsymbol{b}_i | \chi\subseteq B\}].
\eeq
Finally, the unit effect is $\boldsymbol{u}=\sum_{i\in B} \boldsymbol{b}_i$.

 At this point it is straightforward to show that the mathematical structures defining a simplicial GPT are in one-to-one correspondence with the structures defining an ontological theory.

First, one identifies one ontic state $\lambda \in \Lambda$ with each vertex $i \in \mathsf{Vert}[\Delta_{d}]$; that is, one makes a one-to-one correspondence $\lambda \in \Lambda \leftrightarrow i \in \mathsf{Vert}[\Delta_{d}]$.

Next, one can identify the GPT vector space $V$ with the vector space $\mathds{R}^{\Lambda}$ of functions from $\Lambda$ to $\mathds{R}$, where the function associated with each GPT vector is simply its coordinates along each element of the chosen basis; that is, there is a one-to-one correspondence $\mathds{R}^{\Lambda} \leftrightarrow V$.
Explicitly, any vector $\boldsymbol{w} \in V$ is uniquely identified with a function $f_{\boldsymbol{w}} \in \mathds{R}^{\mathsf{Vert}[\Delta_{d}]}$ by   $f_{\boldsymbol{w}}(i) := \left<\boldsymbol{w},\boldsymbol{b}_i\right>_V$,  where $\boldsymbol{b}_i$ is the orthonormal basis used in the definition of the canonical form (it is the set of vectors that point to the vertices of $\Delta_{d}$).
 Clearly, then, the vertices of $\Delta_{d}$ will be represented by point distributions in $\mathds{R}^{\mathsf{Vert}[\Delta_{d}]}$,  as $f_{\boldsymbol{b}_j}(i) = \left<\boldsymbol{b}_j,\boldsymbol{b}_i\right>_V  = \delta_{ij}$.  By linearity, the other vectors in $\Delta_{d}$ will correspond to mixtures of these, and hence to probability distributions over  $\mathsf{Vert}[\Delta_{d}]$. As such, there is a one-to-one correspondence $\mathcal{D}[\Lambda] \leftrightarrow \Delta_{d}$.

 Similarly, the vectors in $\Delta_{d}^*$ can be identified with the response functions.   The extremal points of $\Delta_{d}^*$, namely, $\textstyle\sum_{i\in \chi}\boldsymbol{b}_i$, are represented by the functions: $f_{\scaleto{\textstyle\sum_{j\in \chi}\boldsymbol{b}_j}{8pt}}(i) = \textstyle\sum_{j\in \chi} \left<\boldsymbol{b}_j,\boldsymbol{b}_i\right>_V = \sum_{j\in \chi} \delta_{ij}$, and these correspond to the extremal response functions in $\mathcal{F}[\Lambda]$. Again by linearity, we see that every vector in $\Delta_{d}^*$ is identified with a valid response function. Additionally, the vector ${\bf u}$ corresponds to a vector ${\bf 1}$ whose components are all equal to $1$. Hence, we have the one-to-one correspondences $F[\Lambda] \leftrightarrow \Delta_{d}^*$ and ${\bf 1} \leftrightarrow {\bf u}$.

 Finally, the dot product can now be written out in terms of these functions as
 $\left<\boldsymbol{w},\boldsymbol{w}'\right>_V = \sum_{i\in \mathsf{Vert}[\Delta_{d}]} f_{\boldsymbol{w}}(i)f_{\boldsymbol{w}'}(i)$, where the right-hand side defines the dot product in the ontological space. We denote this correspondence as
 $\sum_{\lambda\in\Lambda}f_{\_}(\lambda) f_{\_}(\lambda) \leftrightarrow \left<\_,\_\right>_V$.

 We can summarize all this in a table:
\begin{center}
\begin{tabular}{c|c} \label{table1}
ontological & geometric \\ \hline
    $\lambda \in \Lambda$  &  $i \in \mathsf{Vert}[\Delta_{d}]$  \\
  $\mathds{R}^{\Lambda}$ & $V$ \\
  $\mathcal{D}[\Lambda]$ & $\Delta_{d}$ \\
  $\mathcal{F}[\Lambda]$ & $\Delta_{d}^*$ \\
    $\boldsymbol{1}$ & $\boldsymbol{u}$\\
   $\sum_{\lambda\in\Lambda}f_{\_}(\lambda) f_{\_}(\lambda)$ \blk & $\left<\_,\_\right>_V$
\end{tabular}
\end{center}

Hence, any simplex-embeddable GPT $G$ can be embedded into a simplicial GPT, whose canonical form together with the identification in the table can be used to explicitly construct an ontological model of $G$.

Conversely,  any GPT which admits of an ontological model
 can be embedded into a simplex and its dual, which can be explicitly constructed using the identifications provided in the table.

This establishes the theorem.\endproof

\section{Example of a dimension mismatch: Stabilizer rebit}\label{app:dimensionmismatch}

We now give an example of a simplex-embeddable GPT whose state and effect spaces cannot be embedded into a simplex and its dual (respectively) {\em in a vector space of the same dimension as the GPT itself}. The example is the GPT associated to the well-known subtheory of the stabilizer qubit theory containing only the real-amplitude states and effects, termed the {\em stabilizer rebit} quantum subtheory.

 For the sake of obtaining a GPT representation of this theory, we will conceptualize the projectors as vectors in the real vector space of Hermitian operators (as opposed to the complex vector space of arbitrary operators). The GPT state space of the stabilizer rebit is then the convex subset
\begin{equation}
\mathsf{ConvexHull}\left[\left\{\ketbra{0}{0},\ketbra{1}{1},\ketbra{+}{+},\ketbra{-}{-}\right\}\right]
\end{equation}
of the full qubit state space:
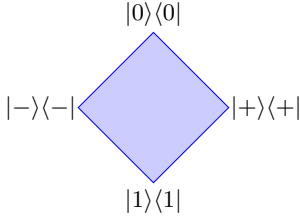
\[
\begin{tikzpicture}
	\begin{pgfonlayer}{nodelayer}
		\node [style=none] (0) at (0, 2) {};
		\node [style=none] (1) at (-2, -0) {};
		\node [style=none] (2) at (0, -2) {};
		\node [style=none] (3) at (2, -0) {};
		\node [style=none] (4) at (0, 2.5) {\small$\ketbra{0}{0}$};
		\node [style=none] (5) at (0, -2.5) {\small$\ketbra{1}{1}$};
		\node [style=none] (6) at (-3, -0) {\small$\ketbra{-}{-}$};
		\node [style=none] (7) at (3, -0) {\small$\ketbra{+}{+}$};
	\end{pgfonlayer}
	\begin{pgfonlayer}{edgelayer}
		\filldraw[blue!20,draw=blue] (1.center) to (0.center) to (3.center) to (2.center) to cycle;
	\end{pgfonlayer}
\end{tikzpicture}}
\]
and the effect space is the convex subset
\begin{equation}
\mathsf{ConvexHull}[\{0,\ketbra{0}{0},\ketbra{1}{1},\ketbra{+}{+},\ketbra{-}{-},\mathds{1}\}]
\end{equation}
of the full qubit effect space.

This stabilizer rebit GPT admits of an ontological model; one such model is the toy model of Ref.~\cite{spekkens2007evidence}.  (We present this model explicitly at the end of this appendix.) As demanded by Theorem~\ref{thm:Simplex-NC}, this implies that the stabilizer rebit state space can be embedded in a simplex whose dual contains the stabilizer rebit effect space. Indeed, the space of probability distributions over ontic states
 of the toy model of Ref.~\cite{spekkens2007evidence} defines a tetrahedron which contains the state space, and whose dual contains the effect space.
Note that the stabilizer rebit state space is $2$-dimensional, while the simplex (the tetrahedron) in which it is embedded is intrinsically $3$-dimensional. As we now prove, this dimension mismatch is unavoidable: it is not possible to find an embedding into any intrinsically $2$-dimensional simplex such that the effect space embeds in the dual.

\proof
We assume that there exists a triangle which embeds the stabilizer rebit state space, whose dual embeds the stabilizer rebit effect space, and we then prove a contradiction.

Note that for any of the extremal states, there exists an effect that evaluates to zero uniquely on that state, for example given the state $\ketbra{0}{0}$ then we have the effect $\ketbra{1}{1}$ that gives probability zero only for that state. The set of states for which this effect evaluates to zero, i.e. those satisfying $\mathrm{tr}\left[\ketbra{1}{1}\rho\right]=\bra{1}\rho\ket{1}=0$, can be geometrically represented by a hyperplane in the state space:
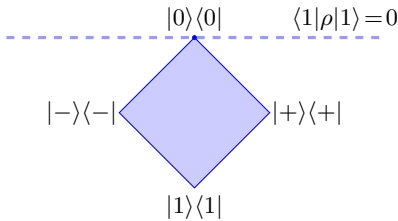
\[
\begin{tikzpicture}
	\begin{pgfonlayer}{nodelayer}
		\node [style=blue dot,scale=0.25] (0) at (0, 2) {};
		\node [style=none] (1) at (-2, -0) {};
		\node [style=none] (2) at (0, -2) {};
		\node [style=none] (3) at (2, -0) {};
		\node [style=none] (4) at (0, 2.5) {\small$\ketbra{0}{0}$};
		\node [style=none] (5) at (0, -2.5) {\small$\ketbra{1}{1}$};
		\node [style=none] (6) at (-3, -0) {\small$\ketbra{-}{-}$};
		\node [style=none] (7) at (3, -0) {\small$\ketbra{+}{+}$};
		\node [style=none] (8) at (-5, 2) {};
		\node [style=none] (9) at (5, 2) {};
		\node [style=none] (10) at (4, 2.5) {\small$\bra{1}\rho\ket{1}=0$};
	\end{pgfonlayer}
	\begin{pgfonlayer}{edgelayer}
		\filldraw[blue!20,draw=blue] (1.center) to (0.center) to (3.center) to (2.center) to cycle;
		\draw [style={thick blue dashed edge}] (8.center) to (9.center);
	\end{pgfonlayer}
\end{tikzpicture}}
\]
Let us consider the embedding of the state $\ketbra{0}{0}$ into the triangle. A priori this could go to any point in the triangle. However, we know that there should be some effect in the dual that evaluates to zero on this state, and not on any of the other states we will embed. This means that $\ketbra{0}{0}$ must be mapped to a point on the boundary of the triangle. (The only effect that evaluates to zero on an interior point of the triangle is the zero effect which would also evaluate to zero on all of the other states).

This means that $\ketbra{0}{0}$ must lie on a proper face of the triangle, and the associated effect defines the hyperplane that picks out that face. There are two possibilities here, $\ketbra{0}{0}$ is mapped to a vertex, or an edge, that is
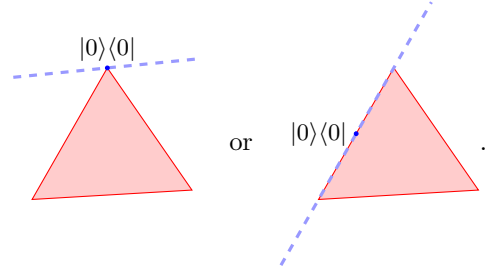
\[
\begin{tikzpicture}
	\begin{pgfonlayer}{nodelayer}
		\node [style={blue dot}, scale=0.25] (0) at (0, 2) {};
		\node [style=none] (1) at (-2, -1.5) {};
		\node [style=none] (2) at (2.25, -1.25) {};
		\node [style=none] (3) at (2.5, 2.25) {};
		\node [style=none] (4) at (-2.5, 1.75) {};
		\node [style=none] (5) at (0, 2.5) {\small$\ketbra{0}{0}$};
	\end{pgfonlayer}
	\begin{pgfonlayer}{edgelayer}
		\filldraw[red!20,draw=red] (1.center) to (0.center) to (2.center) to cycle;
		\draw [style={thick blue dashed edge}] (4.center) to (3.center);
	\end{pgfonlayer}
\end{tikzpicture}}\quad\text{or}\quad %
\begin{tikzpicture}
	\begin{pgfonlayer}{nodelayer}
		\node [style=none] (0) at (0, 2) {};
		\node [style=none] (1) at (-2, -1.5) {};
		\node [style=none] (2) at (2.25, -1.25) {};
		\node [style=none] (3) at (1, 3.75) {};
		\node [style=none] (4) at (-3, -3.25) {};
		\node [style=none] (5) at (-2, 0.25) {\small$\ketbra{0}{0}$};
		\node [style={blue dot}, scale=0.25] (6) at (-1, 0.2500001) {};
	\end{pgfonlayer}
	\begin{pgfonlayer}{edgelayer}
		\filldraw[red!20,draw=red] (1.center) to (0.center) to (2.center) to cycle;
		\draw [style={thick blue dashed edge}] (4.center) to (3.center);
	\end{pgfonlayer}
\end{tikzpicture}}\text{.}
\]

Now consider a second state, e.g. $\ketbra{+}{+}$. The same argument applies again, so it must be mapped to some proper face of the triangle. Moreover, this must be disjoint from the face into which we embedded $\ketbra{0}{0}$, since otherwise the effect associated to $\ketbra{0}{0}$ would also evaluate to zero on $\ketbra{+}{+}$, for example:
 
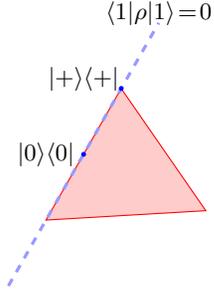
\[
\begin{tikzpicture}
	\begin{pgfonlayer}{nodelayer}
		\node [style={blue dot}, scale=0.25] (0) at (0, 2) {};
		\node [style=none] (1) at (-2, -1.5) {};
		\node [style=none] (2) at (2.25, -1.25) {};
		\node [style=none] (3) at (1, 3.75) {};
		\node [style=none] (4) at (-3, -3.25) {};
		\node [style=none] (5) at (-2, 0.25) {\small$\ketbra{0}{0}$};
		\node [style={blue dot}, scale=0.25] (6) at (-1, 0.2500001) {};
		\node [style=none] (7) at (-1, 2.25) {\small$\ketbra{+}{+}$};
		\node [style=none] (8) at (1, 4) {\small$\bra{1}\rho\ket{1}=0$};
	\end{pgfonlayer}
	\begin{pgfonlayer}{edgelayer}
		\filldraw[red!20,draw=red] (1.center) to (0.center) to (2.center) to cycle;
		\draw [style={thick blue dashed edge}] (4.center) to (3.center);
	\end{pgfonlayer}
\end{tikzpicture}}
\]
That is, it would imply that $\mathrm{tr}\left[\ketbra{1}{1}\ketbra{+}{+}\right]=0$, and so the ontological model would give predictions different from those of the GPT. Hence, $\ketbra{0}{0}$ and $\ketbra{+}{+}$ must be mapped to disjoint faces, for example:
 
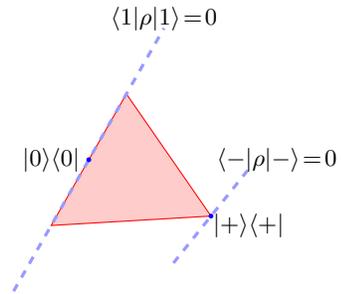
\[
\begin{tikzpicture}
	\begin{pgfonlayer}{nodelayer}
		\node [style=none] (0) at (0, 2) {};
		\node [style=none] (1) at (-2, -1.5) {};
		\node [style=none] (2) at (2.25, -1.25) {};
		\node [style=none] (3) at (1, 3.75) {};
		\node [style=none] (4) at (-3, -3.25) {};
		\node [style=none] (5) at (-2, 0.25) {\small$\ketbra{0}{0}$};
		\node [style={blue dot}, scale=0.25] (6) at (-1, 0.2500001) {};
		\node [style=none] (7) at (3.25, -1.5) {\small$\ketbra{+}{+}$};
		\node [style=none] (8) at (1, 4) {\small$\bra{1}\rho\ket{1}=0$};
		\node [style={blue dot}, scale=0.25] (9) at (2.25, -1.25) {};
		\node [style=none] (10) at (1.249999, -2.5) {};
		\node [style=none] (11) at (4, 0.2500001) {\small$\bra{-}\rho\ket{-}=0$};
		\node [style=none] (12) at (3.25, -0) {};
	\end{pgfonlayer}
	\begin{pgfonlayer}{edgelayer}
		\filldraw[red!20,draw=red] (1.center) to (0.center) to (2.center) to cycle;
		\draw [style={thick blue dashed edge}] (4.center) to (3.center);
		\draw [style={thick blue dashed edge}] (10.center) to (12.center);
	\end{pgfonlayer}
\end{tikzpicture}}
\]
We can apply this same reasoning to every pair of the four extremal states of the stabilizer rebit
 to conclude that these states must all be mapped to disjoint faces of the triangle. 
 But there are at most three disjoint faces of a triangle, so we reach a contradiction.
\endproof

\subsubsection{Explicit ontological model for the stabilizer rebit}
This ontological model is obtained directly by taking the appropriate subsets of the state and effect spaces from the toy theory of Ref.~\cite{spekkens2007evidence}.
The ontic state space is defined as
\beq
\Lambda\equiv\{\lambda_{0+},\lambda_{0-},\lambda_{1+},\lambda_{1-}\}.
\eeq
The epistemic states are given by 
\begin{eqnarray}
\tilde{\mu}_{\ketbra{0}{0}}&=&(1/2,1/2,0,0),\nonumber\\
\tilde{\mu}_{\ketbra{1}{1}}&=&(0,0, 1/2,1/2),\nonumber\\
\tilde{\mu}_{\ketbra{+}{+}}&=&(1/2,0,1/2,0),\nonumber\\
\tilde{\mu}_{\ketbra{-}{-}}&=&(0,1/2,0,1/2).
\end{eqnarray}
The response functions are given by
\begin{eqnarray}
\tilde{\xi}_{\ketbra{0}{0}}=(1,1,0,0),\nonumber\\
\tilde{\xi}_{\ketbra{1}{1}}=(0,0,1,1),\nonumber\\
\tilde{\xi}_{\ketbra{+}{+}}=(1,0,1,0),\nonumber\\
\tilde{\xi}_{\ketbra{-}{-}}=(0,1,0,1).
\end{eqnarray}
It is straightforward to verify  that this model reproduces the operational predictions.

\section{Comparing simplex embeddability and simpliciality}

It is interesting to consider under which conditions the notions of simplex embeddability and simpliciality coincide---that is, when the notions of weak and strong non-classicality coincide.
 Intuitively, this is connected to the no-restriction hypothesis, which states that every logically possible effect is a valid physical effect and similarly that every logically possible state is a valid physical state. This is because if a GPT is simplex-embeddable but not itself simplicial, then the extra vectors in the simplex (or its dual) constitute logically but not physically possible states (effects), and so the GPT cannot satisfy the no-restriction hypothesis.


 If one assumes no dimension gap (i.e., the simplex embedding is into an inner product space of the same dimension as the GPT vector space),
then this intuition about the connection between the no-restriction hypothesis and weak and strong notions of classicality can be proved in a particularly simple manner.  \blk



\begin{theoremNew}
If a  GPT $G'=(V,\left<\_,\_\right>_V,\Omega', \mathcal{E}')$ satisfies the no-restriction hypothesis (that is, $\mathcal{E}'=\Omega'^*$) and is moreover simplex-embeddable using an inner product space $(W,\left<\_,\_\right>_W)$ of the same dimension as $V$,
then $G'$ is simplicial.
\end{theoremNew}
\proof
Consider the simplex embedding of  a GPT $G'=(V,\left<\_,\_\right>_V,\Omega', \mathcal{E}')$ using an inner product space $(W,\left<\_,\_\right>_W)$ via linear maps $\iota$ and $\kappa$, such that $\iota(\Omega')\subseteq \Delta$ and $\kappa(\mathcal{E}')\subseteq \Delta^*$. As we are assuming that the dimension of $W$ is the same as the dimension of $V$, it is clear that the GPT $G:=(W,\left<\_,\_\right>_W,\iota(\Omega'),\kappa(\mathcal{E}'))$ is equivalent to $G'$ in the sense of Def.~\ref{equivGPT}.  For the remainder of this proof, therefore, we will work with this equivalent GPT, $G$, and notate $\Omega:=\iota(\Omega')$ and $\mathcal{E}:=\kappa(\mathcal{E}')$.
 Since $G'$ satisfies the no-restriction hypothesis (by assumption), so does $G$.



By the assumption of simplex embeddability, the set $\Omega$ of all states of $G$, is contained within a simplex, $\Omega \subseteq \Delta$, and consequently every effect in the hypercube that is dual to this simplex, $\Delta^*$, gives valid probabilities on all states in $\Omega$. It follows that every effect in $\Delta^*$ is a logically possible effect.  Under the no-restriction hypothesis, therefore, it follows that every such effect must be physically possible, that is, $\Delta^* \subseteq \mathcal{E}$.  But the assumption that $G$ is simplex-embeddable also implies that every physically realizable effect is contained in the hypercube $\Delta^*$, so that $\mathcal{E} \subseteq \Delta^*$. The conjunction of these gives that $\mathcal{E}=\Delta^*$. By a final appeal to the no restriction hypothesis, we have that the state space is the dual of the effect space, $\Omega = \mathcal{E}^*$, and consequently $\Omega=\Delta$.

\endproof

Unlike the stronger result proven below, this theorem holds only for embeddings into simplicial GPTs of the same dimension as the given GPT.

This lack of a dimension gap is what was assumed in related work~\cite{Shahandeh} by Shahandeh, which considered the consequences of a notion termed {\em broad noncontextuality}.  The latter differs from the notion of generalized noncontextuality insofar as the set of response functions associated to physically realizable GPT effects are assumed to be tomographically complete not just for distributions associated to physically realizable GPT states over (which is all that generalized noncontextuality demands) but for all distributions over ontic states. This additional assumption implies that any simplex embedding will be without a dimension gap.

Secondly, in Ref.~\cite{Schmid2020}, it was demonstrated that the lack of a dimension gap can be {\em derived} under certain assumptions (in particular, the assumption that the GPT is tomographically local).

However, the above theorem can be strengthened to remove the assumption that there is no dimension gap. The following is an adaptation of a theorem first proved by Shahandeh in \cite{Shahandeh}.

\begin{theoremNew}[Classicality of GPTs satisfying the no-restriction hypothesis]\label{secondtheorem}
For any GPT which satisfies the no-restriction hypothesis, that is, where $\mathcal{E}=\st^*$, simplex-embeddability and simpliciality are equivalent. 
\end{theoremNew}

\proof[Proof]
 Using Theorem~\ref{thm:Simplex-NC} we know that simplex embeddability is equivalent to the existence of an ontological model of the GPT.
This ontological model is specified in terms of some linear representation maps $\tilde{\mu}_\_$ and $\tilde{\xi}_\_$.
 The fact that $\tilde{\mu}_\_$ is a linear map on $\Omega$ implies that, for each $\lambda$, $\tilde{\mu}_{\_}(\lambda)$ is a linear functional on $\Omega$, and the fact that $\tilde{\xi}_\_$ is a linear map on $\mathcal{E}$ implies that, for each $\lambda$, $\tilde{\xi}_{\_}(\lambda)$ is a linear functional on $\mathcal{E}$.
 Riesz's representation theorem further implies that for each $\lambda$ there must exist vectors ${\mathbf e}_\lambda$ and $\mathbf{s}_\lambda$ in $V$ such that
\begin{align}
\tilde{\mu}_{\_}(\lambda) &= \left<{\bf e}_\lambda , \_ \right> \\
\tilde{\xi}_{\_}(\lambda) &= \left<\_,{\bf s}_\lambda \right>.
\end{align}
The no-restriction hypothesis then implies that ${\bf e}_\lambda \in \mathcal{E}$ and ${\bf s}_\lambda\in \Omega$ for all $\lambda \in \Lambda$.

Next, note that for any fixed $\mathbf s \in \Omega$, it holds that for all $\mathbf e \in \mathcal{E}$:
\begin{align}
\left<\mathbf{e},\mathbf{s}\right> & = \sum_{\lambda} \tilde{\mu}_{\bf s}(\lambda)\tilde{\xi}_{\mathbf e}(\lambda)\\
&=\sum_{\lambda} \tilde{\mu}_{\bf s}(\lambda) \left<\mathbf{e},{\bf s}_\lambda \right> \\
&=\left< \mathbf{e}, \sum_{\lambda}\tilde{\mu}_{\bf s}(\lambda) {\bf s}_{\lambda}\right>,
\end{align}
and hence tomography of the GPT implies that
\beq \label{convdecomps}
\mathbf{s} = \sum_{\lambda} \tilde{\mu}_{\bf s}(\lambda) {\bf s}_{\lambda}
\eeq
for every $\mathbf{s}\in\Omega$.
%
Because $\tilde{\mu}_{\bf s}(\lambda)$ is a probability distribution, this implies that every GPT state vector $\mathbf s$ can be written as a convex combination of the states $\{{\bf s}_\lambda\}$.

Now consider some vertex $\mathbf{v} \in \mathsf{Vert}[\Omega]$, and decompose it as
\beq
\mathbf{v} = \sum_\lambda \tilde{\mu}_\mathbf{v}(\lambda)\mathbf{s}_\lambda.
\eeq
As $\mathbf{v}$ is a vertex (i.e., it is convexly extremal), it must be the case that for any $\lambda$ such that $\tilde{\mu}_\mathbf{v}(\lambda)\neq 0$, we have $\mathbf{s}_\lambda = \mathbf{v}$.
 This immediately implies that distinct vertices have ontic representations with disjoint support on the ontic state space.  The proof is by contradiction.  Consider distinct vertices $\mathbf{v},\mathbf{w}\in\mathsf{Vert}[\Omega]$ and assume that their representations have overlapping support, that is, assume that there exists $\lambda^* \in\ \Lambda$ such that $\tilde{\mu}_\mathbf{v}(\lambda^*)\neq 0$ and $\tilde{\mu}_\mathbf{w}(\lambda^*) \neq  0$.  In this case, we could infer that
 $\mathbf{v}=\mathbf{s}_{\lambda^*}$ and $\mathbf{w}=\mathbf{s}_{\lambda^*}$, contradicting the hypothesis that $\mathbf{v}$ and $\mathbf{w}$ are distinct. 


Now let us consider two decompositions of some state $\mathbf{s}^* \in \Omega$ into vertices:
\beq
\sum_{\mathbf{v}\in \mathsf{Vert}[\Omega]} \alpha_\mathbf{v} \mathbf{v} = \mathbf{s}^* = \sum_{\mathbf{w}\in \mathsf{Vert}[\Omega]} \beta_\mathbf{w} \mathbf{w}
\eeq
If the state space is not a simplex, then there {\em must} exist non-unique decompositions. However, we will now show that these two decompositions are necessarily the same, and hence that $\Omega$ is a simplex.

Applying $\tilde{\mu}$ to these two decompositions, linearity of $\tilde{\mu}$ implies that
\beq
\sum_{\mathbf{v}\in \mathsf{Vert}[\Omega]} \alpha_\mathbf{v} \tilde{\mu}_\mathbf{v} = \sum_{\mathbf{w}\in \mathsf{Vert}[\Omega]} \beta_\mathbf{w} \tilde{\mu}_\mathbf{w}.
\eeq
Since the $\tilde{\mu}_\mathbf{v}$ all have disjoint support, it must be that $\alpha_\mathbf{v}=\beta_\mathbf{v}$ for all $\mathbf{v}\in \mathsf{Vert}[\Omega]$, and hence the two decompositions are in fact identical. So there are no non-unique decompositions of states, and hence $\Omega$ must be a simplex. By the assumed no-restriction hypothesis, we also infer that $\mathcal{E} = \Omega^*$, and hence the GPT is simplicial.

\endproof

What this means is that any GPT that is simplex-embeddable but not itself
simplicial asserts the existence of some candidate states or some
candidate effects (or some of both) which are logically possible (in
the sense that they would yield valid probabilities) but
stipulated to be not physically realizable. That is, such GPTs must necessarily violate the no-restriction hypothesis.
The divergence of weak and strong notions of nonclassicality is therefore only manifest when we go beyond the special class of GPTs which satisfy the no-restriction hypothesis.

To show that the distinction between weak and strong nonclassicality disappears for all {\em and only} GPTs satisfying the no-restriction hypothesis, it remains only to establish the `only' half of the implication.  This is the easy half.  The only GPTs 
which 
satisfy {\em both} the condition of being simplicial and the condition of being simplex-embeddable are those
 that are simplicial, and these clearly satisfy the no-restriction hypothesis.

\end{document}